\documentclass[iop]{emulateapj}

\usepackage{natbib,aas_macros,amsmath}
\usepackage{multirow,color}
\usepackage{natbib}

\newcommand{\carcsec}{$\!\!\arcsec$}
\newcommand{\m}[1]{\mathrm{#1}}

\newcommand{\redc}[1]{\textcolor{black}{#1}}
\newcommand{\redcc}[1]{\textcolor{black}{#1}}

\begin{document}
\shortauthors{Harikane et al.}
\slugcomment{Accepted for publication in ApJ}

\shorttitle{
High {\sc [Oiii]}/{\sc [Cii]} Ratios in High Redshift Galaxies
}

\title{
Large Population of ALMA Galaxies at $z>6$\\
with Very High {\sc [Oiii]}$88\mathrm{\mu m}$ to {\sc [Cii]}$158\mathrm{\mu m}$ Flux Ratios:\\
Evidence of Extremely High Ionization Parameter or PDR Deficit?
}

\email{yuichi.harikane@nao.ac.jp}
\author{
Yuichi Harikane\altaffilmark{1,2},
Masami Ouchi\altaffilmark{1,3,4},
Akio K. Inoue\altaffilmark{5,6},
Yoshiki Matsuoka\altaffilmark{7},
Yoichi Tamura\altaffilmark{8},
Tom Bakx\altaffilmark{8,1},
Seiji Fujimoto\altaffilmark{6,1},
Kana Moriwaki\altaffilmark{9},
Yoshiaki Ono\altaffilmark{3},
Tohru Nagao\altaffilmark{7},
Ken-ichi Tadaki\altaffilmark{1},
Takashi Kojima\altaffilmark{3,9},
Takatoshi Shibuya\altaffilmark{10},
Eiichi Egami\altaffilmark{11},
Andrea Ferrara\altaffilmark{12},
Simona Gallerani\altaffilmark{12},
Takuya Hashimoto\altaffilmark{6},
Kotaro Kohno\altaffilmark{13},
Yuichi Matsuda\altaffilmark{1,14},
Hiroshi Matsuo\altaffilmark{1,14},\\
Andrea Pallottini\altaffilmark{15,12},
Yuma Sugahara\altaffilmark{3,9,6,1},
Livia Vallini\altaffilmark{16}
}

\affil{$^1$
National Astronomical Observatory of Japan, 2-21-1 Osawa, Mitaka, Tokyo 181-8588, Japan
}
\affil{$^2$
Department of Physics and Astronomy, University College London, Gower Street, London WC1E 6BT, UK
}
\affil{$^3$
Institute for Cosmic Ray Research, The University of Tokyo, 5-1-5 Kashiwanoha, Kashiwa, Chiba 277-8582, Japan
}
\affil{$^4$
Kavli Institute for the Physics and Mathematics of the Universe (WPI), University of Tokyo, Kashiwa 277-8583, Japan
}
\affil{$^5$
Department of Physics, School of Advanced Science and Engineering, Waseda University, 3-4-1 Okubo, Shinjuku, Tokyo 169-8555, Japan
}
\affil{$^6$
Waseda Research Institute for Science and Engineering, 3-4-1 Okubo, Shinjuku, Tokyo 169-8555, Japan
}
\affil{$^7$
Research Center for Space and Cosmic Evolution, Ehime University, Bunkyo-cho, Matsuyama, Ehime 790-8577, Japan
}
\affil{$^8$
Division of Particle and Astrophysical Science, Graduate School of Science, Nagoya University, Nagoya 464-8602, Japan
}
\affil{$^9$
Department of Physics, Graduate School of Science, The University of Tokyo, 7-3-1 Hongo, Bunkyo, Tokyo, 113-0033, Japan
}
\affil{$^{10}$
Kitami Institute of Technology, 165 Koen-cho, Kitami, Hokkaido 090-8507, Japan
}
\affil{$^{11}$
Steward Observatory, University of Arizona, 933 N. Cherry Ave., Tucson, AZ 85721, USA
}
\affil{$^{12}$
Scuola Normale Superiore, Piazza dei Cavalieri 7, 50126 Pisa, Italy
}
\affil{$^{13}$
Institute of Astronomy, Graduate School of Science, The University of Tokyo, 2-21-1 Osawa, Mitaka, Tokyo 181-0015, Japan
}
\affil{$^{14}$
Department of Astronomical Science, The Graduate University for Advanced
Studies (SOKENDAI), 2-21-1 Osawa,Mitaka, Tokyo 181-8588, Japan
}
\affil{$^{15}$
Centro Fermi, Museo Storico della Fisica e Centro Studi e Ricerche "Enrico Fermi", Piazza del Viminale 1, Roma, 00184, Italy
}
\affil{$^{16}$
Leiden Observatory, Leiden University, PO Box 9500, 2300 RA Leiden, The Netherlands
}

\begin{abstract}
We present our new ALMA observations targeting {\sc[Oiii]}88$\mu$m, {\sc[Cii]}158$\mu$m, {\sc[Nii]}122$\mu$m, and dust continuum emission for three Lyman break galaxies at $z=6.0293-6.2037$ identified in the Subaru/Hyper Suprime-Cam survey.
We clearly detect {\sc[Oiii]} and {\sc[Cii]} lines from all of the galaxies at $4.3-11.8\sigma$ levels, and identify multi-band dust continuum emission in two of the three galaxies, allowing us to estimate infrared luminosities and dust temperatures simultaneously.
In conjunction with previous ALMA observations for six galaxies at $z>6$, we confirm that all the nine $z=6-9$ galaxies have high {\sc[Oiii]}/{\sc[Cii]} ratios of $L_\m{[OIII]}/L_\m{[CII]}\sim3-20$, $\sim10$ times higher than $z\sim0$ galaxies.
We also find a positive correlation between the {\sc[Oiii]}/{\sc[Cii]} ratio and the Ly$\alpha$ equivalent width (EW) at the $\sim90\%$ significance level.
We carefully investigate physical origins of the high {\sc[Oiii]}/{\sc[Cii]} ratios at $z=6-9$ using Cloudy, and find that high density of the interstellar medium, low C/O abundance ratio, and the cosmic microwave background attenuation are responsible to only a part of the $z=6-9$ galaxies.
Instead, the observed high {\sc[Oiii]}/{\sc[Cii]} ratios are explained by $10-100$ times higher ionization parameters or low photodissociation region (PDR) covering fractions of $0-10\%$, both of which are consistent with our {\sc[Nii]} observations.
The latter scenario can be reproduced with a density bounded nebula with PDR deficit, which would enhance the Ly$\alpha$, Lyman continuum, and $\mathrm{C^+}$ ionizing photons escape from galaxies, consistent with the {\sc[Oiii]}/{\sc[Cii]}-Ly$\alpha$ EW correlation we find.
\end{abstract}


\section{Introduction}\label{ss_intro}

Understanding properties of the interstellar medium (ISM) is important for galaxy formation.
The metallicity in the gas-phase (hereafter metallicity) can be a tracer of the past star formation history including gas inflow and outflow, because heavy elements produced in stars through star formation activities are returned into the ISM \citep[e,g.,][]{2019A&ARv..27....3M}.
The ionization parameter, $q_\m{ion}$, is also an important quantity that characterizes the ionization state in the ISM, defined by
\begin{equation}
q_\m{ion}=\frac{Q_0}{4\pi R^2_\m{s} n_\m{H}},
\end{equation}
in unit of $\m{cm\ s^{-1}}$, where $Q_0$ and $n_\m{H}$ are the hydrogen ionizing photon production rate and hydrogen density, respectively.
$R_\m{s}$ is the Str\"{o}mgren radius defined by
\begin{equation}
Q_\m{0}=\frac{4}{3}R^3_\m{s}n^2_\m{H}\alpha_\m{B}\epsilon,
\end{equation}
where $\alpha_\m{B}$ and $\epsilon$ are the case B recombination rate and the volume filling factor, respectively.
The ionization parameter is sometimes normalized by the light speed $c$,
\begin{equation}
U_\m{ion}=\frac{q_\m{ion}}{c}.
\end{equation}
The ionization parameter can be also related to the escape fraction of ionizing photons that is key for cosmic reionization physics \citep[e.g.,][]{2014MNRAS.442..900N}.

Nebular emission lines are powerful tools to investigate the ISM properties of galaxies.
Spectroscopic observations for rest-frame optical lines reveal that galaxies at $z\sim2-3$ have lower metallicities \citep[e.g.,][]{2008A&A...488..463M}, higher ionization parameters \citep[e.g.,][]{2014MNRAS.442..900N}, and higher densities \citep[e.g.,][]{2015MNRAS.451.1284S,2016ApJ...816...23S,2017ApJ...835...88K}, compared to local galaxies.
However, it is difficult to study the ISM properties of higher redshift galaxies, especially at $z>5$, because commonly-used rest-frame optical emission lines, such as H$\alpha$, H$\beta$, {\sc[Oiii]}$\lambda\lambda$4959,5007, and {\sc[Oii]}$\lambda\lambda$3726,3729 are redshifted out from the atmospheric window in the near-infrared, and ground-based telescopes cannot observe these lines.
Although rest-frame ultraviolet (UV) metal lines such as {\sc[Ciii]}1907, {\sc Ciii]}1909 are used to study $z\sim6-8$ galaxies \citep[e.g.,][]{2015MNRAS.454.1393S,2015MNRAS.450.1846S,2017MNRAS.464..469S}, these lines are typically weak and not always detected \citep[e.g.,][]{2018PASJ...70S..15S,2018MNRAS.479.1180M}.
Broad-band photometry with the {\it Spitzer} Space Telescope gives some constraints on the rest-frame optical lines at $z>4$ \citep[e.g.,][]{2016ApJ...821..122F,2016ApJ...823..143R,2018ApJ...859...84H}, but this method can be used only for galaxies in limited redshift ranges.
Spectroscopy for strong nebular lines is important for understanding the ISM properties of high redshift galaxies, but we need to wait for the launch of the James Webb Space Telescope for the rest-frame optical lines of $z>5$ galaxies.

Far-infrared (FIR) emission lines studied with Atacama Large Millimeter/Submillimeter Array (ALMA) are alternative tools to study the ISM of high redshift galaxies.
Previous ALMA observations report surprisingly weak [{\sc Cii}]158$\mu\mathrm{m}$ emission in Ly$\alpha$ emitters (LAEs) at $z\sim6-7$ compared to local galaxies with similar star formation rates (SFRs), known as the {\sc [Cii]} deficit (e.g., \citealt{2014ApJ...792...34O,2015A&A...574A..19S,2016MNRAS.462L...6K}).
On the other hand, some Lyman break galaxies (LBGs) detected by ALMA at $z\sim5-7$ have [{\sc Cii}] luminosity to SFR ratios ($L_{[\m{CII}]}/SFR$) comparable to local galaxies \citep[e.g.,][]{2015Natur.522..455C,2016ApJ...829L..11P}.
This difference of the $L_{[\m{CII}]}/SFR$ ratio may include key information of an evolution of the ISM properties from low redshift to high redshift.
Based on these observations, \citet{2018ApJ...859...84H} have identified an anti-correlation between $L_\mathrm{[CII]}/SFR$ and the Ly$\alpha$ equivalent width (EW), $\mathrm{EW^0_{Ly\alpha}}$, which is also reported in \citet{2018MNRAS.478.1170C} and \citet{2019ApJ...881..124M}.
Since the Ly$\alpha$ EW well correlates with the Ly$\alpha$ photon escape fraction \citep[e.g.,][]{2017MNRAS.466.1242S,2018ApJ...859...84H,2019A&A...623A.157S}, this anti-correlation indicates that the strength of the {\sc[Cii]} emission is related to the Ly$\alpha$ photon escape from galaxies.

The {\sc[Oiii]}88$\mu$m emission line is predicted to be detectable from $z>6$ galaxies with ALMA \citep{2014ApJ...780L..18I}, and indeed detected \citep{2016Sci...352.1559I,2017ApJ...837L..21L,2017A&A...605A..42C,2018Natur.557..392H,2019ApJ...874...27T,2019PASJ...71...71H}, including the most distant emission line galaxy at $z=9.1096$ \citep{2018Natur.557..392H}.
Given the high success rate of ALMA {[\sc Oiii]} observations, the {[\sc Oiii]} line is one of the most useful tracers for ISM properties at $z\gtrsim6$.
Since [{\sc Oiii}] has an ionization potential ($35.1\ \m{eV}$) higher than {[\sc Cii]} ($11.3\ \m{eV}$), the {[{\sc Oiii}]}/{\sc [Cii]} ratio is useful to investigate ionization state.
However, the number of galaxies with both {[\sc Oiii]} and {[\sc Cii]} observations is limited; for LBGs and LAEs currently only six galaxies are reported (\citealt{2016Sci...352.1559I,2017ApJ...837L..21L,2017A&A...605A..42C,2018Natur.557..392H,2019PASJ...71...71H,2019ApJ...874...27T,2019MNRAS.487L..81L,2020MNRAS.493.4294B}).

ALMA observations have also begun to reveal complex nature of dust-obscured star formation in high redshift universe.
Dust obscuration is usually quantified by the IR to UV luminosity ratio ($IRX=L_\m{IR}/L_\m{UV}$) as a function of the UV slope $\beta_\m{UV}$ (IRX-$\beta_\m{UV}$).
Local starburst galaxies show a correlation between IRX and $\beta_\m{UV}$ \citep[e.g.,][]{1999ApJ...521...64M,2000ApJ...533..682C}, which is known as the Calzetti IRX-$\beta_\m{UV}$ relation.
\citet{2012ApJ...755..144T} modified the Calzetti IRX-$\beta_\m{UV}$ relation considering the aperture effect in the photometry in UV.
The Small Magellanic Cloud (SMC) shows lower IRX values than the Calzetti relation, known as the SMC IRX-$\beta_\m{UV}$ relation \citep[e.g.,][]{1998ApJ...508..539P}.
{\it Herschel} and ALMA observations reveal that $z\sim2-5$ galaxies follow the Calzetti or SMC IRX-$\beta_\m{UV}$ curves \citep[e.g.,][]{2017MNRAS.472..483F,2018ApJ...853...56R,2018MNRAS.476.3991M,2018MNRAS.479.4355K}.
On the other hand at $z>5$, some galaxies show IRX values significantly lower than these relations \citep[e.g.,][]{2015Natur.522..455C,2016ApJ...833...72B,2017ApJ...845...41B}, implying significant evolution of dust properties from $z\sim0$ to $z>5$ \citep[see also][]{2010A&A...523A..85G}.
There are several possibilities for the physical origin of the low IRX values in high redshift galaxies.
One of the most important caveats is that dust temperatures are not determined in most of the galaxies at $z>5$, as discussed in several studies \citep[e.g.,][]{2016ApJ...833...72B,2017ApJ...847...21F,2018MNRAS.477..552B}.
Since these galaxies are often observed in one single FIR band, the dust temperature ($T_\m{dust}$) is usually assumed to local galaxy values to derive the IR luminosity.
However, \citet{2017ApJ...847...21F} point out that dust temperatures of high redshift galaxies could be significantly higher, which increases $L_\m{IR}$ up to 0.6 dex with the higher dust temperature by $\Delta T_\m{dust}=+40\ \m{K}$, possibly resolving the tension between the models and observations \citep[see also][]{2019PASJ...71...71H,2019ApJ...874...27T,2019MNRAS.487L..81L}.
Additionally, using cosmological zoom in simulations \citep{2017MNRAS.471.4128P}, \citet{2018MNRAS.477..552B} shows that 20\% of the dust mass can be responsible for up to 80\% of the IR luminosity, because of the high ($\sim$70 K) dust temperature due to the compactness and intense radiation field typical of high redshift galaxies.
Thus we need to simultaneously constrain dust temperatures and IR luminosities with multi-band dust continuum observations.

In this study, we present our new ALMA observations targeting [{\sc Oiii}]88$\mu$m, [{\sc Cii}]158$\mu$m, [{\sc Nii}]122$\mu$m, and multi-band dust continuum emission in three LBGs at $z\sim6$ that are identified in the Subaru/Hyper Suprime-Cam survey \citep{2018PASJ...70S...4A}.
Optical spectroscopic observations have already detected Ly$\alpha$ emission lines in these galaxies \citep{2018PASJ...70S..35M}.
In conjunction with previous ALMA observations for $z>5$ galaxies and Cloudy model calculations, we study ISM properties of the high redshift galaxies.

\begin{deluxetable*}{cccc}
\setlength{\tabcolsep}{0.4cm}
\tablecaption{Summary of Observational Results of Our Targets}
\startdata
\hline
\hline
 & J1211-0118 & J0235-0532 & J0217-0208\\
\hline
R.A. & 12:11:37.112 & 02:35:42.412 & 02:17:21.603\\
decl. & $-$01:18:16.500 & $-$05:32:41.623 & $-$02:08:52.778\\
$M_\m{UV}$ [AB mag] & $-22.8$ & $-22.8$ & $-23.3$\\
$L_\m{UV}$ [$L_\odot$] & $2.7\times10^{11}$ & $2.9\times10^{11}$ & $4.3\times10^{11}$\\
$EW^0_\m{Ly\alpha}$ [$\mathrm{\AA}$] & $6.9\pm0.8$ & $41\pm2$ & $15\pm1$\\
$\beta_\m{UV}$ & $-2.0\pm0.5$ & $-2.6\pm0.6$ & $-0.1\pm0.5$\\
\hline
$z_\m{[OIII]}$ & $6.0295\pm0.0009$ & $6.0906\pm0.0009$ & $6.2044\pm0.0013$\\
$z_\m{[CII]}$ & $6.0291\pm0.0008$ & $6.0894\pm0.0010$ & $6.2033\pm0.0009$\\
$z_\m{sys}$ & $6.0293\pm0.0002$ & $6.0901\pm0.0006$ & $6.2037\pm0.0005$\\
$z_\m{Ly\alpha}$ & $6.0339\pm0.0008$ & $6.0918\pm0.0002$ & $6.2046\pm0.0006$\\
$\Delta v_\m{Ly\alpha}\ [\m{km\ s^{-1}}]$ & $196\pm35$ & $71\pm26$ & $37\pm32$\\
\hline
$[${\sc Oiii}$]$ integrated flux [Jy km s$^{-1}$] & $2.69\pm0.40\ (6.7\sigma)$ & $2.10\pm0.18\ (11.8\sigma)$ & $4.57\pm1.06\ (4.3\sigma)$\\
$[${\sc Cii}$]$ integrated flux [Jy km s$^{-1}$] & $1.42\pm0.15\ (9.5\sigma)$ & $0.43\pm0.07\ (5.9\sigma)$ & $1.36\pm0.20\ (6.7\sigma)$\\
$[${\sc Nii}$]$ integrated flux [Jy km s$^{-1}$] & $<0.66\ (3\sigma)$ & $<0.90\ (3\sigma)$ & $<0.45\ (3\sigma)$\\
$\m{FWHM_{[OIII]}}$ [km s$^{-1}$] & $194\pm123$ & $389\pm117$ & $374\pm162$\\
$\m{FWHM_{[CII]}}$ [km s$^{-1}$] & $170\pm98$ & $270\pm135$ & $316\pm117$\\
\hline
$L_\m{[OIII]}\ [L_\odot]$& $(4.8\pm0.7)\times10^{9}$ & $(3.8\pm0.3)\times10^{9}$ & $(8.5\pm2.0)\times10^{9}$\\
$L_\m{[CII]}\ [L_\odot]$& $(1.4\pm0.1)\times10^{9}$ & $(4.3\pm0.7)\times10^{8}$ & $(1.4\pm0.2)\times10^{9}$\\
$L_\m{[NII]}\ [L_\odot]$& $<8.3\times10^{8}$ & $<1.2\times10^{9}$ & $<6.2\times10^{8}$\\
$L_\m{[OIII]}/L_\m{[CII]}$ & $3.4\pm0.6$ & $8.9\pm1.7$ & $6.0\pm1.7$\\
$L_\m{[OIII]}/L_\m{[NII]}$ & $>5.8$ & $>3.2$ & $>13.8$\\
\hline
$S_{\nu,160}$ [$\mu$Jy] & $220\pm51\ (4.3\sigma)$ & $<101\ (3\sigma)$ & $239\pm79\ (3.0\sigma)$\\
$S_{\nu,120}$ [$\mu$Jy] & $382\pm72\ (5.3\sigma)$ & $<162\ (3\sigma)$ & $310\pm43\ (7.1\sigma)$\\
$S_{\nu,90}$ [$\mu$Jy] & $<826\ (3\sigma)$ & $<394\ (3\sigma)$ & $<606\ (3\sigma)$\\
\hline
$L_\m{IR}$ [$L_\odot$] & $3.2^{+18.7}_{-1.7}\times10^{11}$ & $<2.5\times10^{11}\ (3\sigma)$ & $1.4^{+2.5}_{-0.3}\times10^{11}$\\
$T_\m{dust}$ [K] & $38^{+34}_{-12}$ & $\dots$ & $25^{+19}_{-5}$\\
$M_\m{dust}$ [$M_\odot$] & $3.0^{+10.5}_{-2.3}\times10^{7}$ & $<6.9\times10^{6}\ (3\sigma)$ & $1.9^{+73.5}_{-1.6}\times10^{8}$\\
\hline
$SFR_\m{tot}$ [$M_\odot\ \m{yr^{-1}}$] & $86$ & $54$ & $96$\\
$SFR_\m{UV}$ [$M_\odot\ \m{yr^{-1}}$] & $51$ & $54$ & $80$\\
$SFR_\m{IR}$ [$M_\odot\ \m{yr^{-1}}$] & $34$ & $<27$ & $16$
\enddata
\label{tab_prop}
\end{deluxetable*}

This paper is organized as follows.
In Section \ref{ss_target}, we describe our targets and optical spectroscopy already conducted.
Our ALMA observations are presented in Section \ref{ss_alma}.
We present our results for the FIR emission lines and dust continuum emission in Sections \ref{ss_line} and \ref{ss_dust}, respectively.
In Section \ref{ss_dis}, we discuss our results based on model calculations by Cloudy.
Section \ref{ss_summary} summarizes our findings.
Throughout this paper we use the recent Planck cosmological parameter sets constrained with the temperature power spectrum, temperature-polarization cross spectrum, polarization power spectrum, low-$l$ polarization, CMB lensing, and external data \citep[TT, TE, EE+lowP+lensing+ext result; ][]{2016A&A...594A..13P}:
$\Omega_\m{m}=0.3089$, $\Omega_\Lambda=0.6911$, $\Omega_\m{b}=0.049$, and $h=0.6774$.
We assume a \citet{2003PASP..115..763C} initial mass function (IMF) with lower and
upper mass cutoffs of $0.1\m{M_\odot}$ and $100\m{M_\odot}$, respectively.
All magnitudes are in the AB system \citep{1983ApJ...266..713O}, and are corrected for
Galactic extinction \citep{1998ApJ...500..525S}.

\section{Target Selection}\label{ss_target}
We select targets for our ALMA observations from the Subaru/Hyper Suprime-Cam Subaru strategic program (HSC-SSP) survey datasets \citep{2018PASJ...70S...4A}.
The Subaru/HSC survey is a photometric survey with optical broad band filters $grizy$ and several narrow-band filters.
The HSC survey has three layers, UltraDeep, Deep, and Wide, with different combinations of area and depth (see \citealt{2018PASJ...70S...4A} for details).
LBGs at $z\sim4-7$ are selected from the HSC datasets with the dropout selection technique \citep{2016ApJ...828...26M,2018PASJ...70S..35M,2018ApJS..237....5M,2019arXiv190807910M,2018PASJ...70S..10O,2018PASJ...70S..11H,2018PASJ...70S..12T}, and some of the LBGs are spectroscopically confirmed \citep{2016ApJ...828...26M,2018PASJ...70S..35M,2018ApJS..237....5M,2019arXiv190807910M,2018PASJ...70S..10O}.

As the first step to understand ISM properties of high redshift galaxies, we select three luminous LBGs, J1211-0118, J0235-0532, and J0217-0208 as targets for ALMA observations (Table \ref{tab_prop}).
All targets are spectroscopically confirmed with Ly$\alpha$ in the SHELLQs project \citep{2018PASJ...70S..35M}, and their redshifts, $6.0<z_\m{Ly\alpha}<6.3$, are suitable for ALMA Band 6, 7, and 8 observations targeting {\sc [Cii]}158$\mu$m, {\sc [Nii]}122$\mu$m, and {\sc [Oiii]}88$\mu$m, respectively.
J1211-0118 shows a weak Ly$\alpha$ emission line with a rest-frame Ly$\alpha$ equivalent width (EW) of $EW^0_\m{Ly\alpha}=6.9\pm0.8\ \m{\AA}$.
J0235-0532 and J0217-0208 show strong Ly$\alpha$ emission lines with $EW^0_\m{Ly\alpha}=41\pm2$, and $15\pm1\ \m{\AA}$, respectively.
The broad dynamical range of the Ly$\alpha$ EW makes our targets representative in terms of Ly$\alpha$ emission.
The spectroscopic data also indicate that the targets do not have broad Ly$\alpha$ ($>400\ \m{km\ s^{-1}}$) nor Nv emission lines.
J0235-0532 and J0217-0208 exhibit large Ly$\alpha$ luminosities of $L_\m{Ly\alpha}>10^{43}\ \m{erg\ s^{-1}}$, powerful enough to be associated with AGNs as suggested by \citet{2016ApJ...823...20K} at $z\sim2$, while a spectroscopic study by \citet{2018PASJ...70S..15S} do not find signatures of AGN activitity (e.g., {\sc Civ} emission) in such luminous LAEs at $z\sim6-7$.
NIR spectroscopy is important for understanding possible AGN activities in J0235-0532 and J0217-0208.

\begin{figure*}[t]
 \begin{center}
  \includegraphics[clip,bb=6 6 427 422,width=0.7\hsize]{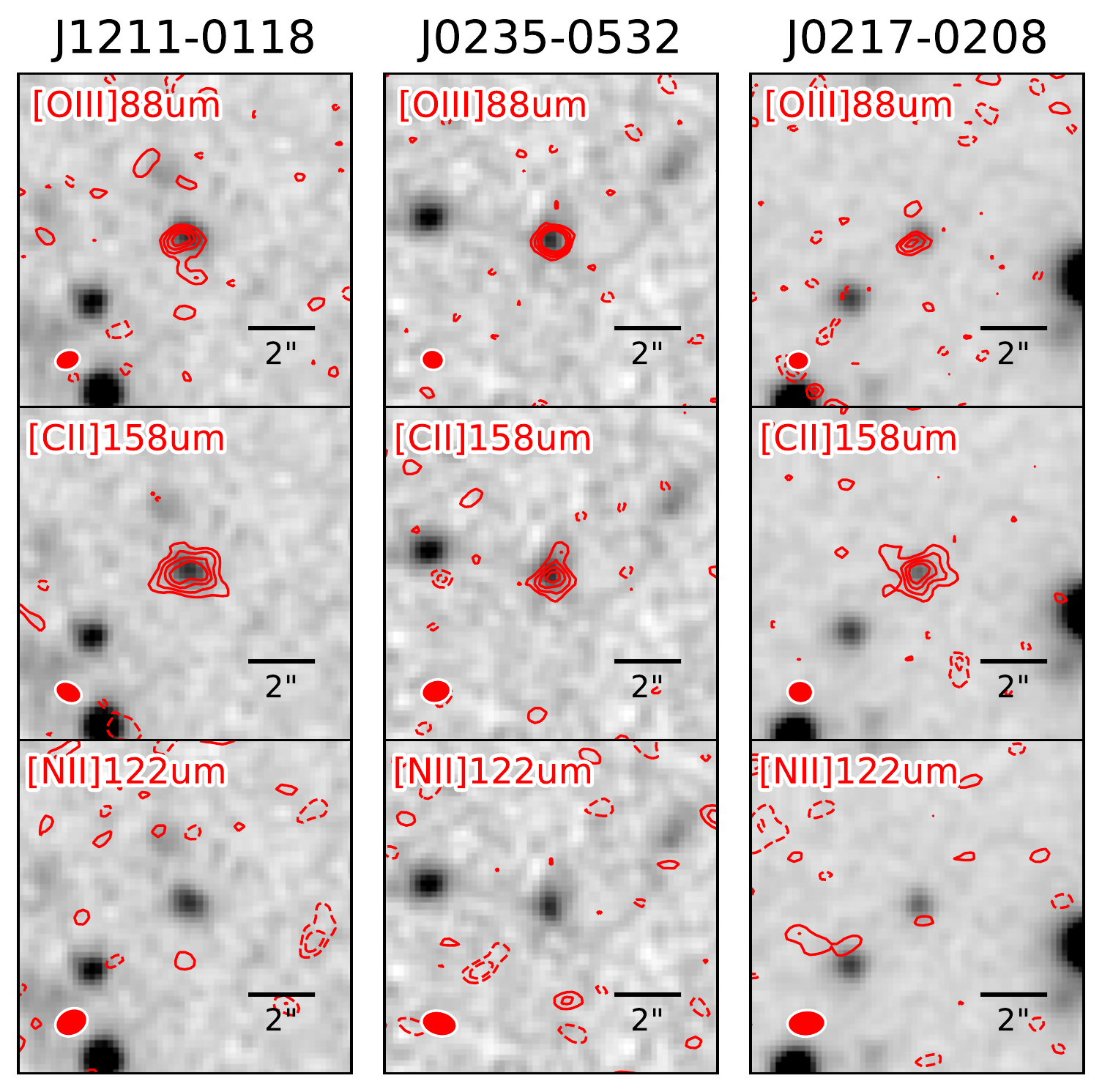}
 \end{center}
   \caption{
{\sc [Oiii]}88$\mu$m, {\sc [Cii]}158$\mu$m, and {\sc [Nii]}122$\mu$m emission of our targets after continuum subtraction.
The red contours are {\sc[Oiii]}, {\sc[Cii]}, and {\sc[Nii]} emission, and are drawn at $1\sigma$ intervals from $\pm2\sigma$ to $\pm5\sigma$.
Positive and negative contours are shown by the solid and dashed lines, respectively.
The backgrounds are rest-UV and Ly$\alpha$ image (the Subaru/HSC $z$-band) whose typical seeing size is 0.\carcsec7.
The images are $10\arcsec\times10\arcsec$, and the red ellipses at the lower left corner indicate the synthesized beam sizes of ALMA.
   \label{fig_snap_line}}
\end{figure*}

\section{ALMA Observations and Data Reduction}\label{ss_alma}
The three LBGs at $z\sim6$ were observed during ALMA cycle 5 (ID: \#2017.1.00508.S, PI: Y. Harikane) at Bands 6, 7, and 8 for [{\sc Cii}]158$\mu$m, [{\sc Nii}]122$\mu$m, and [{\sc Oiii}]88$\mu$m, between 2018 April 3rd and 2018 June 22nd.
The antenna configurations were C43-1, C43-2, and C43-3, achieving the beam sizes of $\sim0.\carcsec6-1.\carcsec1$ and the maximum recoverable scales of $6-7\arcsec$.
We used four spectral windows (SPWs) with 1.875 GHz bandwidths in the Frequency Division Mode and the total band width of 7.5 GHz.
The velocity resolution was set to $\sim3-5\ \m{km\ s^{-1}}$.
One of the SPWs was centered on the [{\sc Cii}], [{\sc Nii}], or [{\sc Oiii}] line frequency expected from the redshift of the Ly$\alpha$ emission.

The data were reduced and calibrated using the Common Astronomy Software \citep[CASA; ][]{2007ASPC..376..127M} pipeline version 5.1.1 in the general manner with scripts provided by the ALMA observatory.
Using the task {\tt CLEAN}, we produced images and cubes with the natural weighting without taper to maximize point-source sensitivities.
To generate a pure dust continuum image, we collapsed all SPWs except for one SPW where the [{\sc Cii}], [{\sc Nii}], or [{\sc Oiii}] line is located, i.e., the continuum is taken at $\gtrsim600\ \m{km\ s^{-1}}$ from the line center.
To create a pure line image, we subtracted continuum using the off-line channels in the line cube with the CASA task {\tt uvcontsub}.

\begin{figure*}[!t]
 \begin{center}
  \includegraphics[clip,bb= 6 6 637 241,width=0.65\hsize]{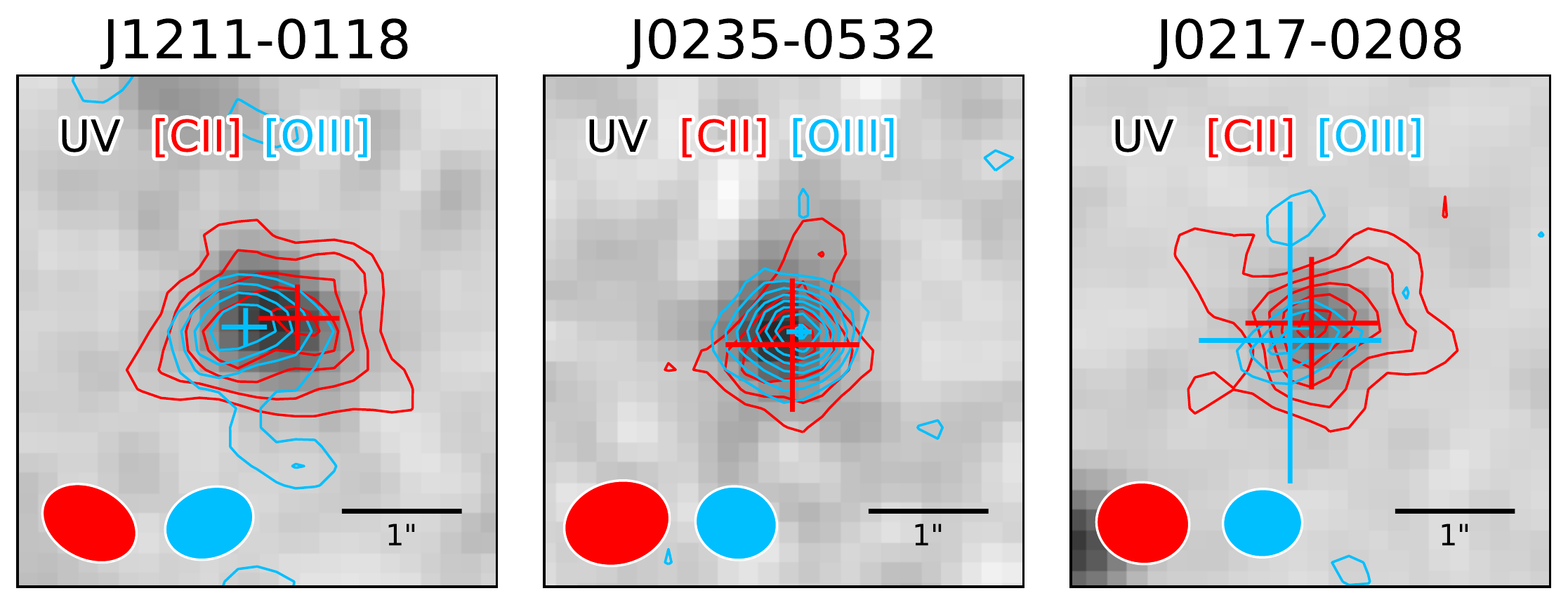}
 \end{center}
   \caption{
ALMA {\sc[Cii]}158$\mu$m and {\sc [Oiii]}88$\mu$m emission maps on rest-UV and Ly$\alpha$ image (the Subaru/HSC $z$-band).
The red and cyan contours show {\sc[Cii]} and {\sc[Oiii]} emission, respectively, and are drawn at $1\sigma$ intervals from $2\sigma$.
The red and cyan crosses show the peak positions, and the sizes show their uncertainties estimated from the Monte Carlo simulations (see text).
   \label{fig_offset_line}}
\end{figure*}

\begin{figure*}[!t]
 \begin{center}
  \includegraphics[clip,bb=7 8 713 409,width=0.8\hsize]{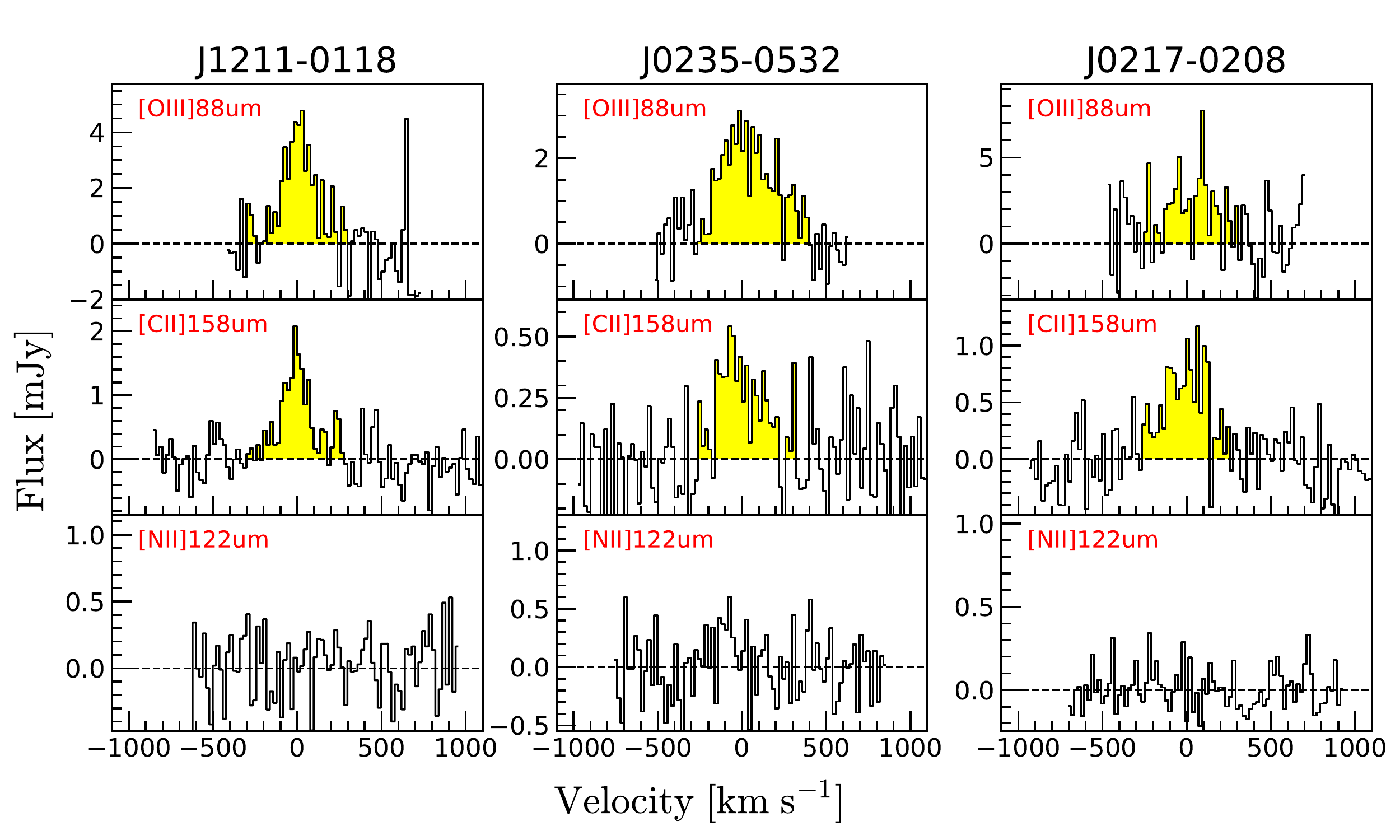}
 \end{center}
   \caption{
ALMA spectra of our targets after the continuum subtraction.
These spectra are extracted from a 0.\carcsec7-diameter circular aperture, re-binned to a spectral resolution of $20\ \m{km\ s^{-1}}$.
The velocity zero point is set to the systemic redshift of each target.
The {\sc[Oiii]} and {\sc[Cii]} emission lines are clearly detected, and their redshifts are consistent within $1\sigma$ uncertainties.
   \label{fig_spec}}
\end{figure*}

\section{FIR Emission Lines}\label{ss_line}

\subsection{Line Detections and Upper limits}

Figure \ref{fig_snap_line} displays the {[\sc Oiii]}88$\mu$m, {\sc [Cii]}158$\mu$m, and {\sc [Nii]}122$\mu$m emission lines of our three targets.
These moment 0 maps were made with the CASA task {\tt immoments}, integrating over $600\ \m{km\ s^{-1}}$ covering most of the velocity range of the line emission ($>1.5\times \m{FWHM}$)
This wavelength range is comparable to previous studies for luminous galaxies \citep{2016Sci...352.1559I,2017ApJ...851..145M,2019PASJ...71...71H}.
We calculate the signal-to-noise ratios (SNRs) of the emission lines using $0.\carcsec7$-diameter circular aperture, by randomly placing apertures on the image and adopting the rms as the noise level.
The {\sc [Oiii]} and {\sc [Cii]} emission lines are clearly detected in all of our targets at the $4.3-11.8\sigma$ significance levels.
In addition, the {\sc [Cii]} emission in J1211-0118 and J0217-0208 are spatially well resolved.
The {\sc [Nii]} emission are not detected in our three targets.
For all three emission lines for the three targets, total fluxes are measured on the $600\ \m{km\ s^{-1}}$-integrated maps in $2\arcsec$-radius circular apertures, which can cover the area where spatially extended emission may exist like the {\sc[Cii]}158$\mu$m halos around $z\sim6$ galaxies reported by \citet{2019arXiv190206760F}.
Note that the maximum recoverable scales of our data, $6-7\arcsec$, are larger than the typical size of the  {\sc[Cii]} halo, $\sim2\arcsec$.

In Figure \ref{fig_offset_line}, we compare spatial distributions of the {\sc [Oiii]} and {\sc[Cii]} with the rest-UV emission including Ly$\alpha$.
We plot the peak pixel positions of the {\sc [Oiii]} and {\sc[Cii]} emission with uncertainties with crosses in Figure \ref{fig_offset_line}.
We estimate the uncertainties of the peak positions using Monte Carlo simulations.
In each dataset of the target and the emission line, we add artificial sky noises to pixels following a Gaussian random distribution with a standard deviation equal to the $1\sigma$ noise of the data, and re-measure the peak positions.
We make a series of 1000 simulations for each dataset, and estimate the uncertainties of the peak positions.
\redc{We find that the peak positions of {\sc [Oiii]} and {\sc[Cii]} agree within $2\sigma$ uncertainties in all of our galaxies.}

\begin{figure*}[!t]
\begin{center}
 \begin{minipage}{0.48\hsize}
 \begin{center}
  \includegraphics[clip,bb=25 15 321 309,width=0.9\hsize]{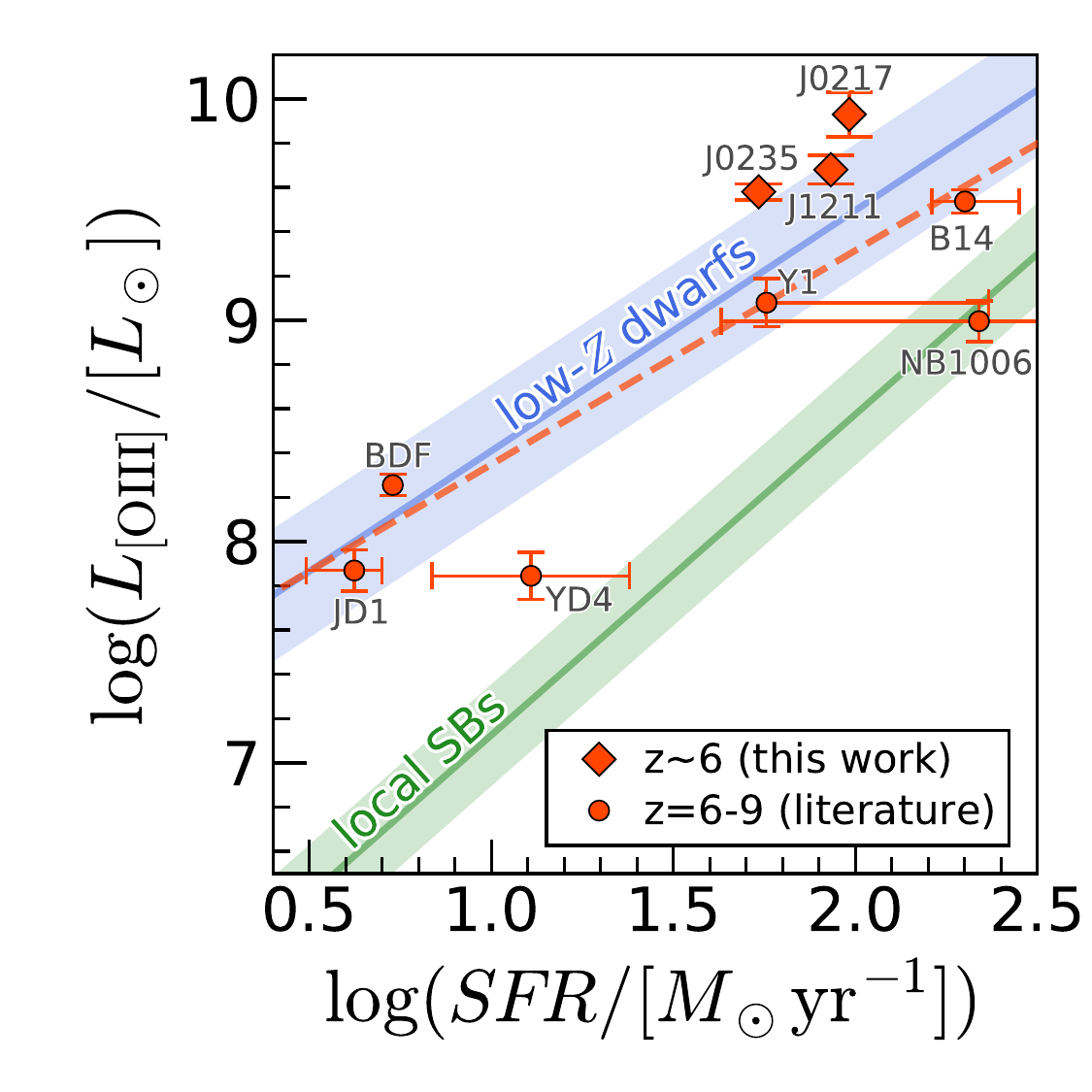}
 \end{center}
 \end{minipage}
  \begin{minipage}{0.48\hsize}
 \begin{center}
  \includegraphics[clip,bb=25 15 321 309,width=0.9\hsize]{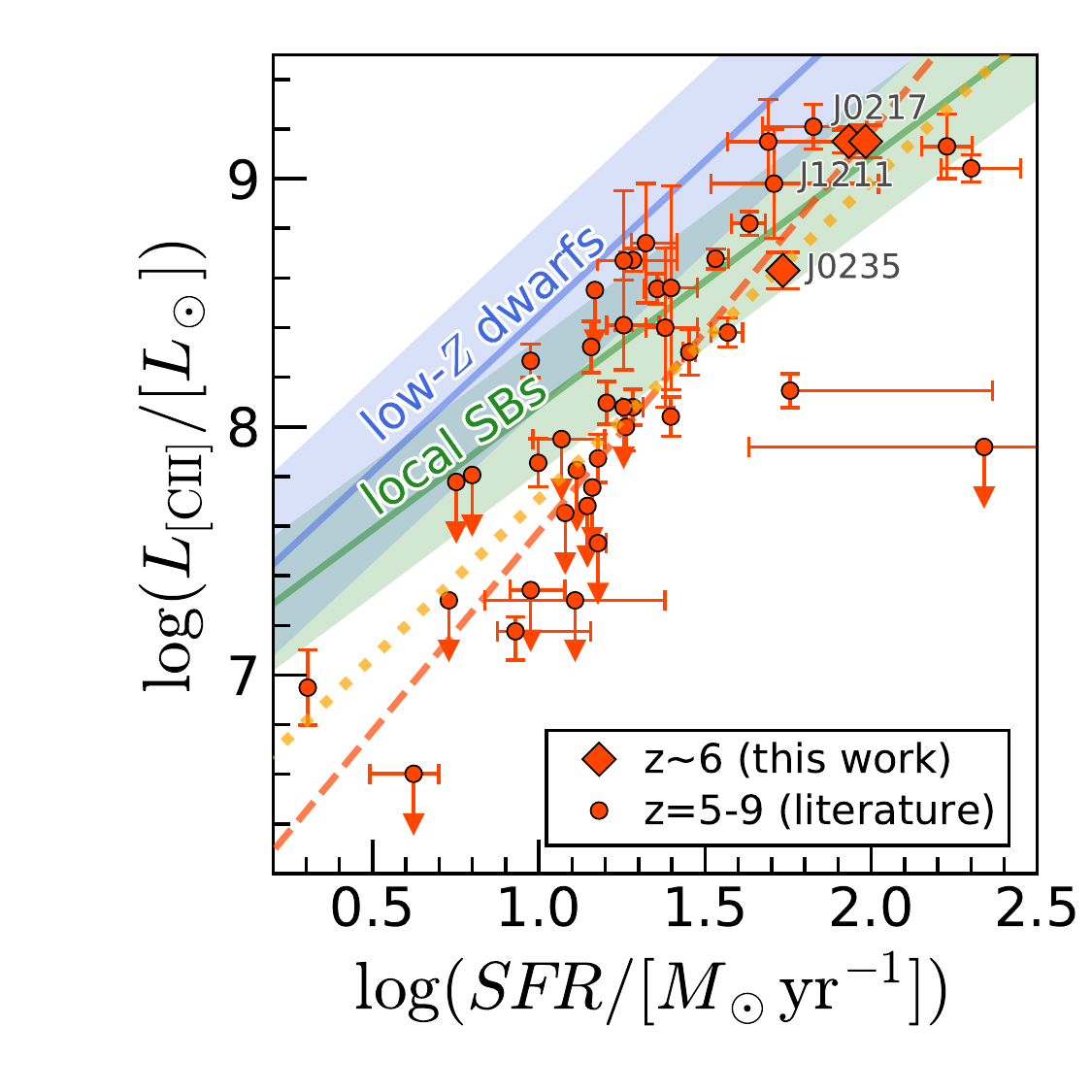}
 \end{center}
 \end{minipage}
 \end{center}
   \caption{
   ({\it Left panel:}) {\sc [Oiii]}88$\mu$m luminosities as a function of the SFR.
    The red diamonds represent our targets at $z\sim6$, and the red circles are other $z=6-9$ LBGs and LAEs in the literature (see Table \ref{tab_lite}). 
    The red dashed line is the fitting function for the $z=6-9$ galaxies (see text).
    The blue and green lines denote relations for $z\sim0$ low-metallicity dwarf galaxies (``low-$Z$ dwarfs'') and starburst galaxies (``local SBs'') from \citet{2014A&A...568A..62D}, respectively.
    The shades correspond to the $1\sigma$ dispersion of the relations.
   ({\it Right panel:}) Same as the left panels but for {\sc [Cii]}158$\mu$m luminosities.
   The red circles are $z=5-9$ LBGs and LAEs in the literature (see Table \ref{tab_lite}).
   \redc{The orange dotted line is a fitting function in \citet{2020arXiv200200979S} for ALPINE sample at $z=4-6$ and $z\sim6-9$ galaxies (their Figure 6).}
   \label{fig_L_SFR}}
\end{figure*}

Figure \ref{fig_spec} shows spectra around the frequencies of the {[\sc Oiii]}88$\mu$m, {\sc [Cii]}158$\mu$m, and {\sc [Nii]}122$\mu$m lines.
We simply fit each spectrum with a single Gaussian profile and the rest-frame {[\sc Oiii]}88$\mu$m, {[\sc Cii]}158$\mu$m, or {[\sc Nii]}122$\mu$m frequency (1900.5369, 3393.0062, or 2459.3801 GHz, respectively\footnote{http://www.cv.nrao.edu/php/splat/}), which delivers the redshift and full width at half maximum (FWHM).
The obtained redshifts and FWHMs are summarized in Table \ref{tab_prop}.
We find that the redshifts based on the {[\sc Oiii]} and {\sc [Cii]} lines are consistent within $1\sigma$ uncertainties.
The FWHMs are $170-390\ \m{km\ s^{-1}}$.
The FWHMs of the {[\sc Oiii]} lines seem to be larger than those of the {\sc [Cii]} lines, but they are comparable within $1\sigma$ uncertainties \citep[see also][]{2019MNRAS.487.1689P}.
We adopt SNR-weighted means of the {[\sc Oiii]} and {\sc [Cii]} redshifts as the systemic redshifts of our targets.
The systemic redshifts of J1211-0118, J0235-0532, and J0217-0208 are $z_\m{sys}=6.0293$, $6.0901$, and $6.2037$, respectively.
The Ly$\alpha$ redshifts measured from the optical spectroscopy are slightly redshifted from the systemic redshifts, because of the resonant scattering of Ly$\alpha$.
The Ly$\alpha$ velocity offsets are $40-200\ \m{km\ s^{-1}}$, comparable to previous results \citep[e.g.,][]{2013ApJ...765...70H,2014ApJ...788...74S,2014ApJ...795...33E,2019PASJ...71...71H}.

Based on the total fluxes and the systemic redshifts, we calculate luminosities of the emission lines.
Table \ref{tab_prop} summarized the calculated {\sc[Oiii]} and {\sc[Cii]} luminosities of and upper limits for {\sc[Nii]}. 
Our targets have some of the highest {\sc[Oiii]} and {\sc[Cii]} luminosities seen in $z>6$ LBGs and LAEs.

\subsection{$L_\m{[OIII]}$-SFR and $L_\m{[CII]}$-SFR Relations}\label{ss_L_SFR}

\begin{figure*}[!t]
 \begin{center}
 \begin{minipage}{0.48\hsize}
 \begin{center}
  \includegraphics[clip,bb=14 15 377 309,width=0.99\hsize]{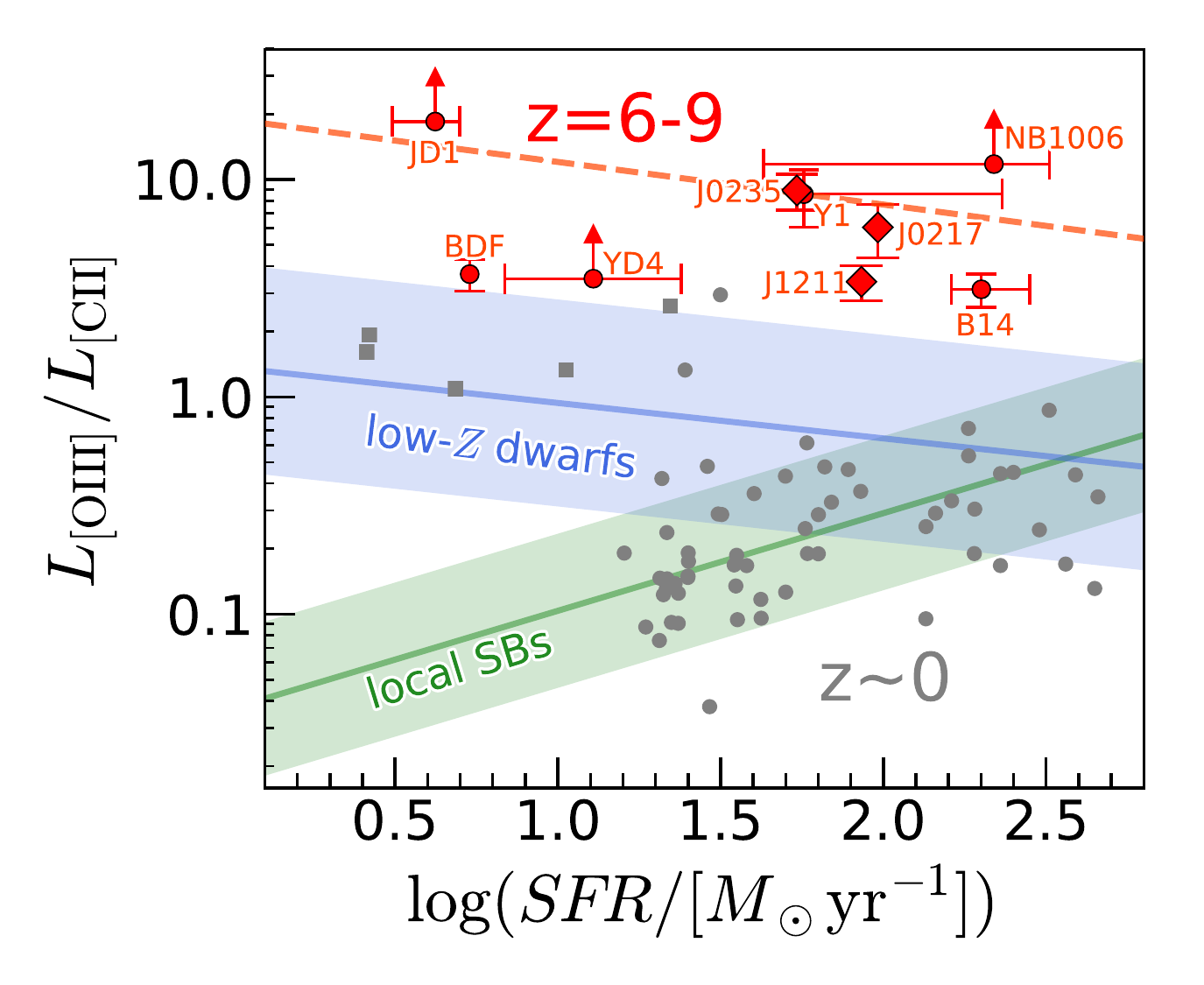}
 \end{center}
 \end{minipage}
  \begin{minipage}{0.48\hsize}
 \begin{center}
  \includegraphics[clip,bb=14 15 377 309,width=0.99\hsize]{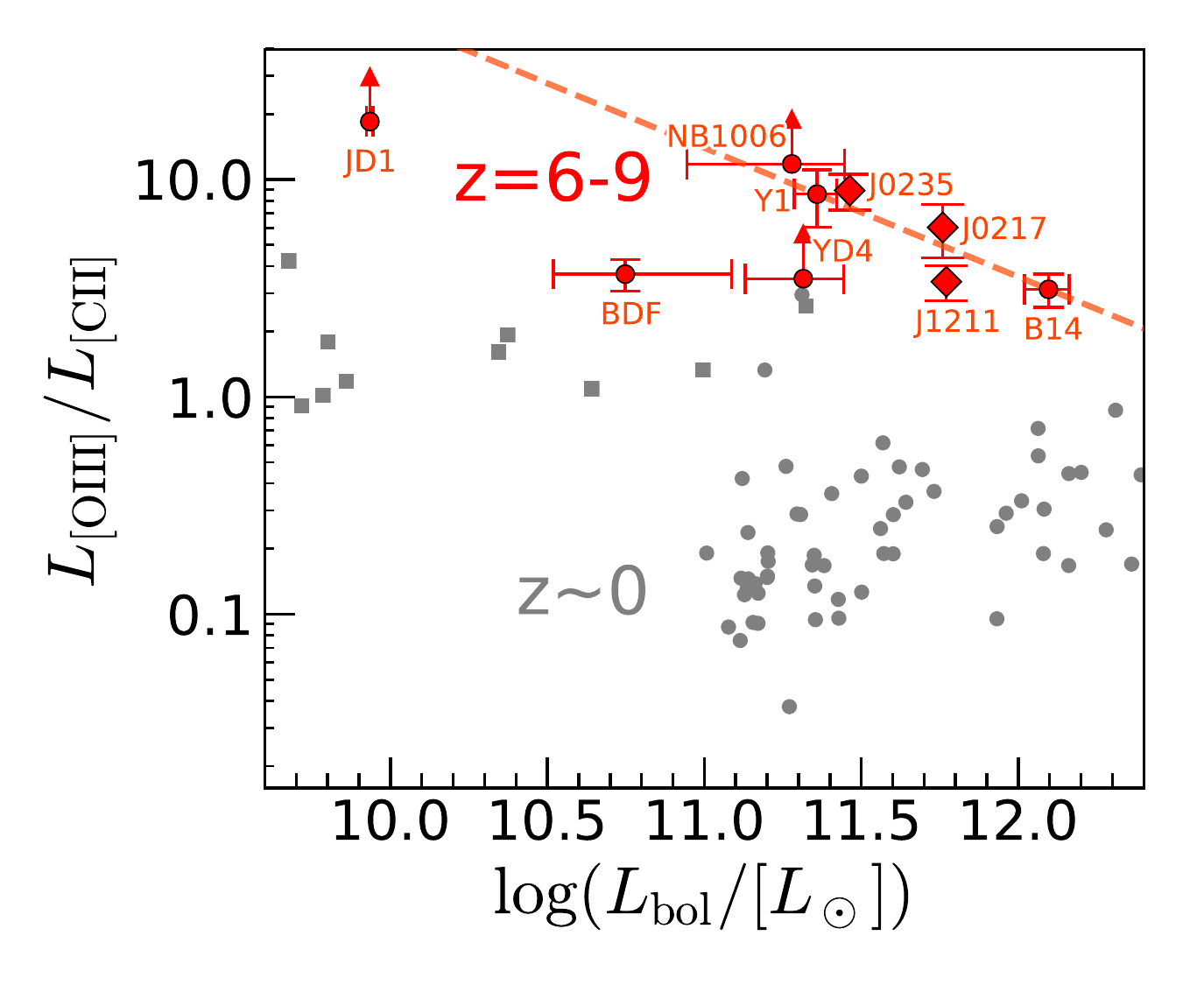}
 \end{center}
 \end{minipage}
 \end{center}
   \caption{
      ({\it Left panel:}) 
{\sc[Oiii]}/{\sc[Cii]} ratios as a function of the SFR.
    The red diamonds represent our targets at $z\sim6$, and the red circles are other $z=6-9$ LBGs and LAEs in the literature (see Table \ref{tab_lite}). 
    The red dashed line is the fitting function for the $z=6-9$ galaxies (see text).
    The gray diamonds and circles denote $z\sim0$ galaxies from  the Dwarf Galaxy Survey \citep{2013PASP..125..600M,2014A&A...568A..62D,2015A&A...578A..53C} and GOALS \citep[][]{2010ApJ...715..572H,2017ApJ...846...32D}, respectively.
    The blue and green lines denote relations for $z\sim0$ low-metallicity dwarf galaxies (``low-$Z$ dwarfs'') and starburst galaxies (``local SBs'') from \citet{2014A&A...568A..62D}, respectively.
    The shades correspond to the $1\sigma$ dispersion of the relations.
    We find that the {\sc[Oiii]}/{\sc[Cii]} ratios of the $z=6-9$ galaxies are systematically higher than those of $z\sim0$ galaxies.
       \redc{({\it Right panel:})
       \redc{
       Same as the left panel but as a function of the bolometric luminosity, $L_\m{bol}$.
       The bolometric luminosity is estimated as a summation of the UV and IR luminosities.
       The UV and IR luminosities of galaxies in the literatures are taken from \citet{2019PASJ...71...71H}.}}
   \label{fig_OIIICII}}
\end{figure*}

\begin{figure*}[!t]
 \begin{center}
  \includegraphics[clip,bb=6 10 506 305,width=0.68\hsize]{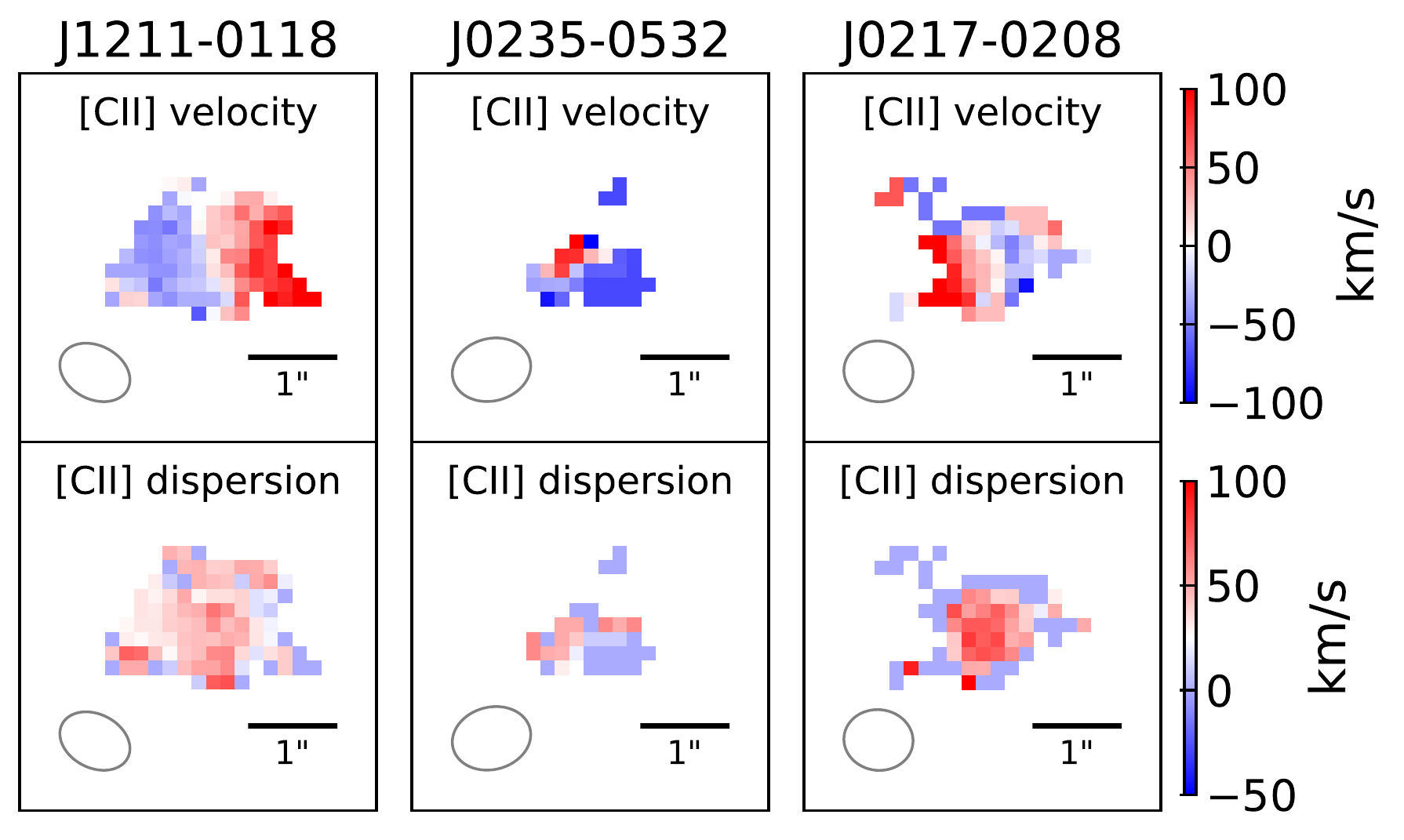}
 \end{center}
   \caption{
Mean velocity map ({\it upper panel}) and velocity dispersion ({\it lower panel}) of {\sc[Cii]} in our targets.
We can identify velocity gradients in J1211-0118 and J0217-0208, suggesting that they are consistent with rotation dominated systems.
   \label{fig_vmap}}
\end{figure*}

We compare {[\sc Oiii]}88$\mu$m and {[\sc Cii]}158$\mu$m luminosities of our targets with previous studies for LBGs and LAEs given SFRs.
We usually plot the {\sc[Cii]} (and {\sc[Oiii]}) luminosity as a function of SFR, because C$^+$ (O$^{2+}$) ionizing photons are made by massive stars recently formed.
The left panel of Figure \ref{fig_L_SFR} shows the {[\sc Oiii]} luminosities as a function of SFR.
For our three targets, these SFRs are total SFRs from UV and IR luminosities that are estimated in Section \ref{ss_LIR_Td}.
We also plot relations for $z\sim0$ local starburst galaxies (local SBs) and low-metallicity dwarf galaxies (low-$Z$ dwarfs) from \citet{2014A&A...568A..62D}, and results of $z=6-9$ galaxies in the literature \citep{2016Sci...352.1559I,2017ApJ...837L..21L,2017A&A...605A..42C,2018Natur.557..392H,2019ApJ...874...27T,2019PASJ...71...71H}, summarized in Table \ref{tab_lite}.
The {\sc[Oiii]} luminosities of our targets are comparable to or higher than the relation for the low-metallicity dwarf galaxies of \citet{2014A&A...568A..62D}.
Other $z=6-9$ galaxies are comparable to either of the local relations.
The $z=6-9$ galaxies do not show any deficit relative to the local $L_\m{[OIII]}$-SFR relations unlike the {\sc[Cii]} case.
In the left panel of Figure \ref{fig_L_SFR}, we plot the following fitting function for the results of $z=6-9$ galaxies as the red dashed line:
\begin{equation}
\m{log}(L_\m{[OIII]}/[L_\odot])=0.97\times\m{log}(SFR/[M_\odot\ \m{yr^{-1}}])+7.4 .
\end{equation}

The right panel of Figure \ref{fig_L_SFR} displays the {\sc [Cii]} luminosities as a function of the SFR.
We also plot relations for $z\sim0$ galaxies and results for $z=5-9$ galaxies in the literature summarized in Table \ref{tab_lite}.
In contrast to the {\sc[Oiii]} luminosities, all of our targets are located below the $L_\m{[CII]}$-SFR relation of $z\sim0$ low-metallicity dwarf galaxies.
We find that J1211-0118 and J0217-0208 follow the relation for $z\sim0$ starburst galaxies, and the {\sc [Cii]} luminosity of J0235-0532 is 0.3 dex lower than that relation.
The relatively low {\sc [Cii]} luminosity of J0235-0532 with the strong Ly$\alpha$ emission ($EW_{Ly\alpha}^0=41\ \m{\AA}$) is consistent with the anti-correlation between $L_\mathrm{[CII]}/SFR$ and $\mathrm{EW^0_{Ly\alpha}}$ reported in \citet{2018ApJ...859...84H}.
The red dashed line in the right panel of Figure \ref{fig_L_SFR} is the fitting function for the results of $z=6-9$ galaxies:
\begin{equation}
\m{log}(L_\m{[CII]}/[L_\odot])=1.6\times\m{log}(SFR/[M_\odot\ \m{yr^{-1}}])+6.0 ,
\end{equation}
which is comparable to a fitting function for a combination of the ALMA Large Program to INvestigate C$^+$ at Early Times (ALPINE) data at $z=4-6$ and $z\sim6-9$ galaxies in \citet[][the orange dotted line in Figure \ref{fig_L_SFR} right]{2020arXiv200200979S}.

\subsection{$L_\m{[OIII]}/L_\m{[CII]}$ Ratios}\label{ss_OIIICII}

\redc{We plot the {\sc [Oiii]}/{\sc [Cii]} luminosity ratio as a function of the SFR and the bolometric luminosities in Figure \ref{fig_OIIICII}, following previous studies \citep{2019MNRAS.487L..81L,2019PASJ...71...71H,2019PASJ...71..109H,2020MNRAS.493.4294B}.}
We also plot results of $z=6-9$ galaxies in the literature and local galaxies studied in the Dwarf Galaxy Survey \citep{2013PASP..125..600M,2014A&A...568A..62D,2015A&A...578A..53C} and the Great Observatories All-sky LIRG Survey \citep[GOALS;][]{2010ApJ...715..572H,2017ApJ...846...32D}.
We find that our targets and other $z=6-9$ galaxies show systematically higher {\sc [Oiii]}/{\sc [Cii]} ratios compared to local galaxies, which is consistent with previous results \citep{2016Sci...352.1559I,2019MNRAS.487L..81L}.
In Figure \ref{fig_OIIICII}, we plot the following fitting functions for the results of $z=6-9$ galaxies as the red dashed lines:
\begin{eqnarray}
\m{log}(L_\m{[OIII]}/L_\m{[CII]})&=&-0.20\times\m{log}(SFR/[M_\odot\ \m{yr^{-1}}])+1.3,\nonumber\\ \\
\m{log}(L_\m{[OIII]}/L_\m{[CII]})&=&-0.59\times\m{log}(L_\m{bol}/[L_\odot])+7.7.
\end{eqnarray}
We will discuss the origin of the high {\sc [Oiii]}/{\sc [Cii]} ratios in Section \ref{ss_dis_OIIICII}.

\subsection{Velocity Fields}

We investigate kinematic properties of our targets using {\sc [Cii]} emission, which are detected with high signal to noise ratios and are spatially well resolved in J1211-0118 and J0217-0208.
With the CASA task {\tt immoments}, we create flux-weighted velocity (i.e., moment 1) maps and velocity dispersion (i.e., moment 2) maps of {\sc [Cii]} emission of our targets.
We only use pixels with $>2\sigma$ detections.
Figure \ref{fig_vmap} shows the velocity and dispersion maps.
The {\sc [Cii]} velocity maps of J1211-0118 clearly shows a $\sim220$ km s$^{-1}$ velocity gradient.
We also identify a gradient of $\sim250$ km s$^{-1}$ in J0217-0208.
The velocity dispersions ($\sigma_\m{tot}\simeq \m{FWHM}/2.35$) are $\sim70$, $\sim110$, and $\sim130$ km s$^{-1}$ for J1211-0118, J0235-0532, and J0217-0208, respectively.
Given the low angular resolution of the observations, there are various interpretations of the velocity gradients.
A rotating galaxy disk would be one interpretation.

\begin{figure*}[!t]
 \begin{center}
  \includegraphics[clip,bb=6 6 427 422,width=0.7\hsize]{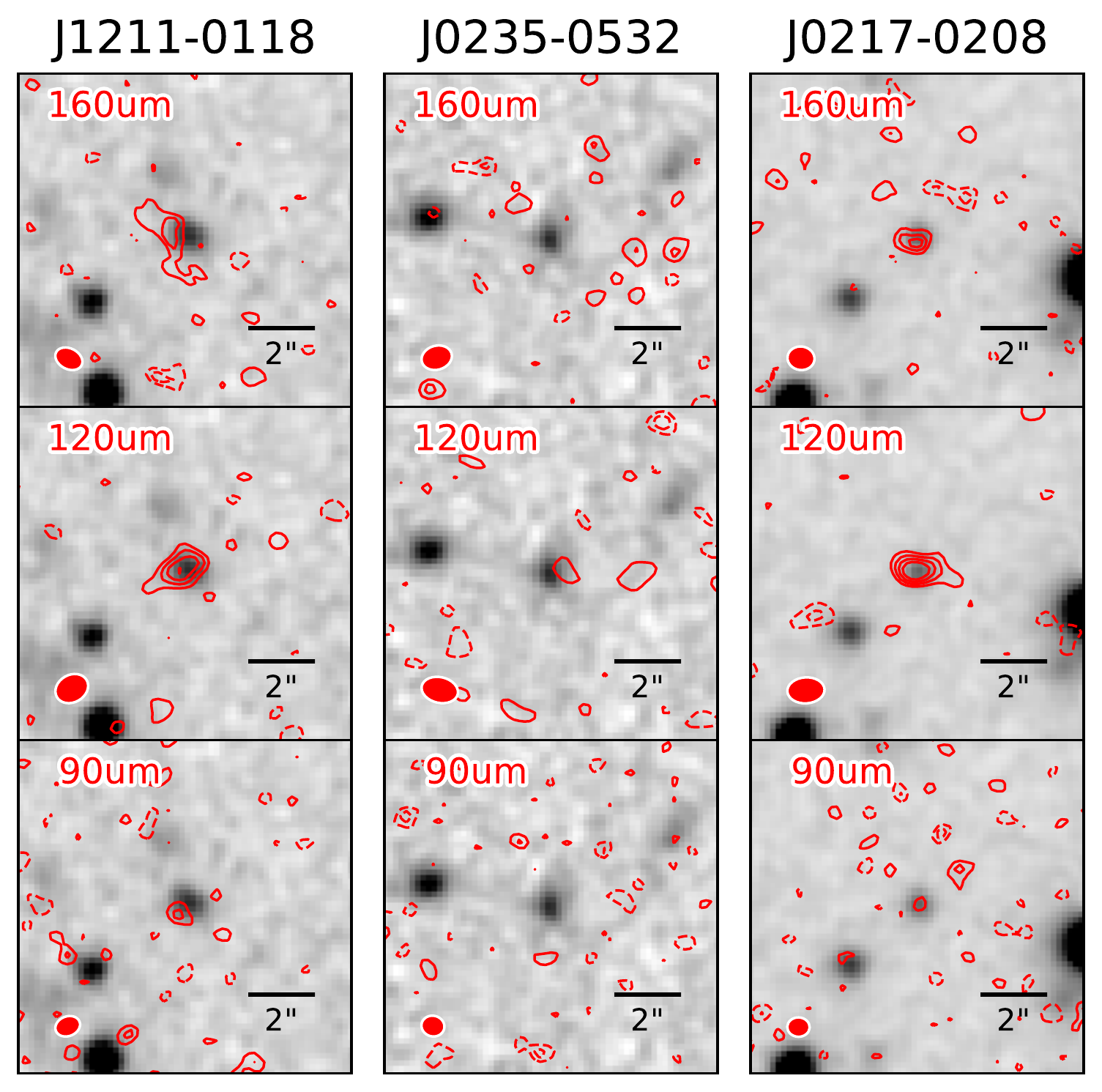}
 \end{center}
   \caption{
Dust continuum emission of our targets.
The red contours are dust emission at $\lambda_\m{rest}=160\mu$m, $120\mu$m, and $90\mu$m, and are drawn at $1\sigma$ intervals from $\pm2\sigma$ to $\pm5\sigma$.
Positive and negative contours are shown by the solid and dashed lines, respectively.
The backgrounds are rest-UV and Ly$\alpha$ image (the Subaru/HSC $z$-band).
The images are $10\arcsec\times10\arcsec$, and the red ellipses at the lower left corner indicate the synthesized beam sizes of ALMA.
   \label{fig_snap_dust}}
\end{figure*}

We apply an observational criterion for the classification of rotation- and dispersion-dominated systems, $\Delta v_\m{obs}/2\sigma_\m{tot}=0.4$, where $\Delta v_\m{obs}$ and $\sigma_\m{tot}$ are the full observed velocity gradient and the velocity dispersions \citep{2009ApJ...706.1364F}.
We find that $\Delta v_\m{obs}/2\sigma_\m{tot}=1.6$ and $0.96$ for J1211-0118 and J0217-0208, respectively, indicating that these two galaxies are consistent with rotation dominated systems.
These values ($\Delta v_\m{obs}/2\sigma_\m{tot}>0.4$) are contrast to those of other systems that are interpreted as galaxy mergers \cite[e.g.,][]{2019ApJ...881L..23B}.

\section{Dust Continua}\label{ss_dust}

\subsection{Detections and Upper limits}

\begin{figure*}[!t]
 \begin{center}
  \includegraphics[clip,bb= 6 6 637 241,width=0.65\hsize]{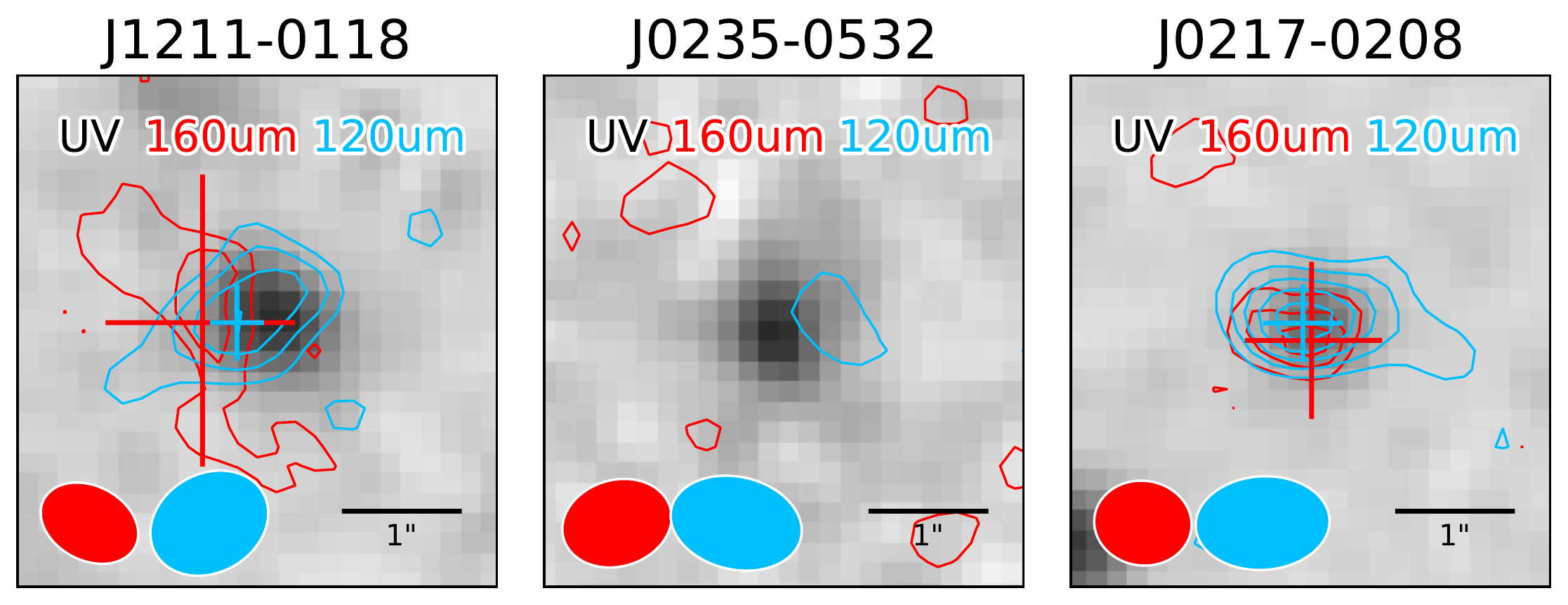}
 \end{center}
   \caption{
ALMA dust emission maps on rest-UV and Ly$\alpha$ image (the Subaru/HSC $z$-band).
The red and cyan contours show dust emission at $\lambda_\m{rest}=160\mu$m and $120\mu$m, respectively, and are drawn at $1\sigma$ intervals from $2\sigma$.
The red and cyan crosses show the peak positions, and the size show their uncertainties estimated from the Monte Carlo simulations.
   \label{fig_offset_dust}}
\end{figure*}

Figure \ref{fig_snap_dust} shows the continuum emission maps of our targets.
We calculate the SNRs of the dust continuum using $0.\carcsec7$-diameter circular aperture, by randomly placing apertures on the image and adopting the rms as the noise level.
We detect dust continuum emission at $\sim120\ \mu\m{m}$ from J1211-0118 and J0217-0208 at $5.3\sigma$ and $7.1\sigma$ significance levels, respectively.
We also identify dust continua of J1211-0118 and J0217-0208 at $\sim160\ \mu\m{m}$, which are useful to constrain dust temperatures of these galaxies (see Section \ref{ss_LIR_Td}).
We do not detect dust continuum emission from J0235-0532.
We compare spatial positions and morphologies of these dust continuum emission with the rest-UV emission in Figure \ref{fig_offset_dust}.
In J1211-0118 and J0217-0208, the peak positions of the dust emission overlap with the rest-UV counterparts, given the uncertainties of the peak determination.

\begin{figure*}
 \begin{center}
  \includegraphics[clip,bb=21 6 636 314,width=0.8\hsize]{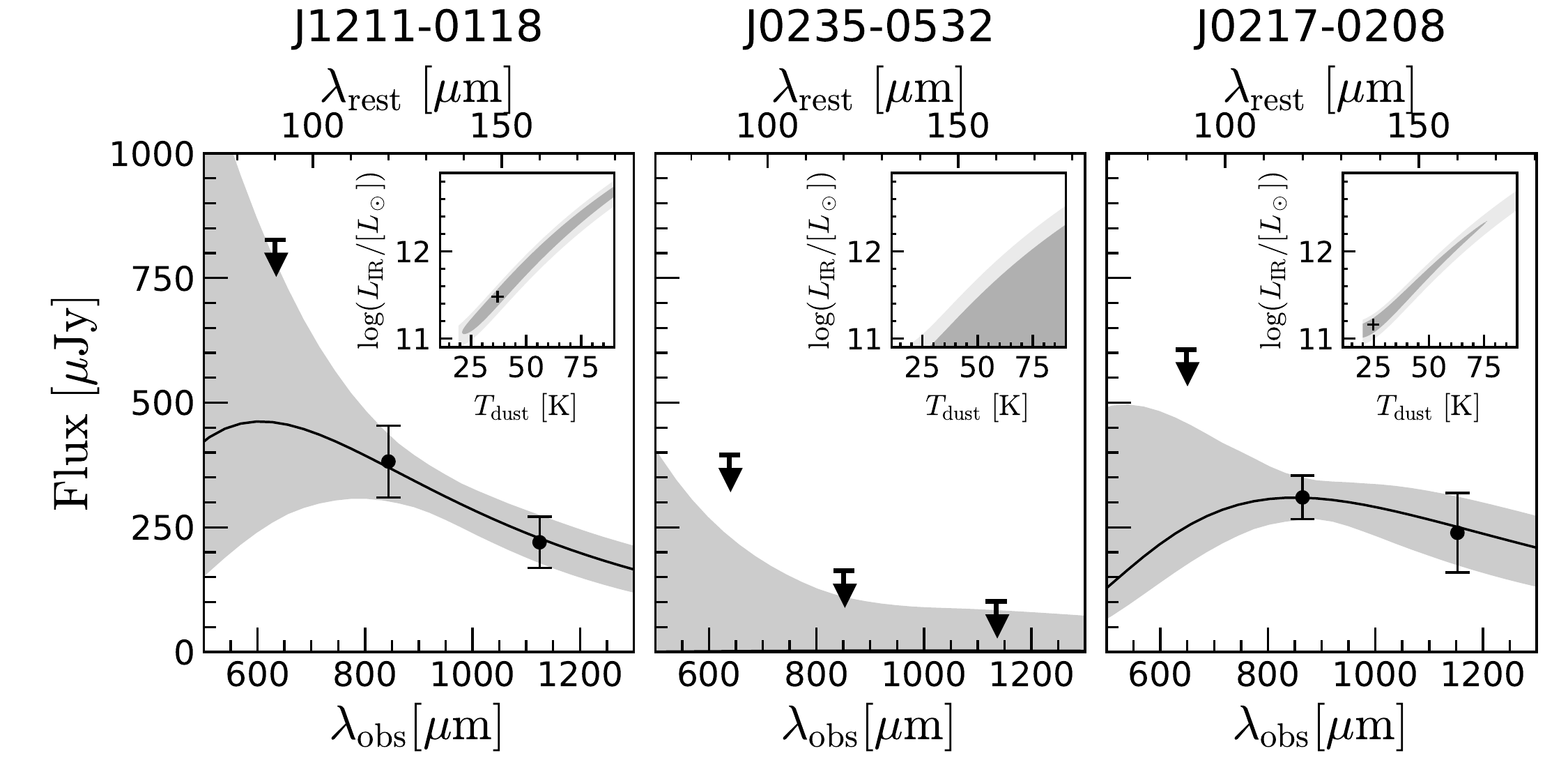}
 \end{center}
   \caption{
Dust SEDs of our targets. 
The black circles and downward arrows are our observations.
The black solid curves and the shade regions show the best-fit modified black body models and the $1\sigma$ uncertainties.
The inset panels show error contours indicating the 68\% and 95\% confidence regions.
The black crosses denote the best-fit parameters.
   \label{fig_fit_mbb}}
\end{figure*}

\subsection{$L_\m{IR}$ and $T_\m{dust}$ Estimates}\label{ss_LIR_Td}

We estimate IR luminosities, $L_\m{IR}$, by integrating the modified black body radiation over $8-1000\ \mu$m.
We fit the observed fluxes with the modified black body by varying the IR luminosity and dust  temperatures, $T_\m{dust}$.
We fix the dust emissivity to $\beta_\m{d}=1.5$, but the effect of this assumption is not significant for our conclusions.
For example, if we change the dust emissivity in $1<\beta_\m{d}<3$, the IR luminosity varies only $<0.1$ dex.
The CMB heating and attenuation effects are included following formulae by \citet{2013ApJ...766...13D}.
\redc{In the fitting, we require that the dust temperature must be higher than the CMB temperature at the redshift of the galaxy ($\sim20\ \m{K}$ at $z\sim6$).}

Figure \ref{fig_fit_mbb} shows the dust spectral energy distributions (SEDs) and the best-fit models for our targets.
We obtain IR luminosities and dust temperatures of $(L_\m{IR}, T_\m{dust})=(3.2^{+18.7}_{-1.7}\times10^{11}\ L_\odot, 38^{+34}_{-12}\ \m{K})$ and ($1.4^{+2.5}_{-0.3}\times10^{11}\ L_\odot, 25^{+19}_{-5}\ \m{K})$, for J1211-0118 and J0217-0208, respectively.
These IR luminosities ($L_\m{IR}>10^{11}\ L_\odot$) correspond to luminous infrared galaxies (LIRGs) in local universe.
These dust temperatures are comparable to other $z>6$ galaxies \citep{2017MNRAS.466..138K,2019PASJ...71...71H} and theoretical simulations \citep{2018MNRAS.477..552B}.
Although the IR luminosity is degenerate with the dust temperature, shown in the error contours in Figure \ref{fig_fit_mbb}, the IR luminosities range in $1.1\times10^{11}<L_\m{IR}/L_\odot<7.9\times10^{12}$ and $1.0\times10^{11}<L_\m{IR}/L_\odot<2.2\times10^{12}$ for J1211-0118 and J0217-0208, respectively, within the $1\sigma$ confidence regions in the error contours.
For J0235-0532, we obtain the $3\sigma$ upper limit of $L_\m{IR}\times T_\m{dust}<1.3\times10^{13}\ L_\odot\m{K}$.
Since we want to compare our observations with previous studies including \citet{2019PASJ...71...71H}, who assume $T_\m{dust}=50\ \m{K}$ for galaxies without continuum detections, we adopt $L_\m{IR}<2.5\times10^{11}\ L_\odot$ assuming $T_\m{dust}=50\ \m{K}$ as the fiducial value for J0235-0532.

\subsection{IRX-$\beta_\m{UV}$ Relation}

We investigate whether our targets follow the known IRX-$\beta$ relations, or show IRX values below the relations.
We estimate the UV slope $\beta$ using the Subaru/HSC photometry and spectroscopic properties of the Ly$\alpha$ emission lines.
Based on the HSC $z$ and $y$ band magnitudes, redshifts, and Ly$\alpha$ EWs, we estimate the UV spectral slope to be $\beta_\m{UV}=-2.0\pm0.5$, $-2.6\pm0.6$, and $-0.1\pm0.5$ for J1211-0118, J0235-0532, and J0217-0208, respectively.
Note that the HSC $z$ and $y$ bands correspond to the rest-frame wavelength of $1300-1500\ \m{\AA}$.

\begin{figure}
 \begin{center}
  \includegraphics[clip,bb=19 21 377 343,width=0.9\hsize]{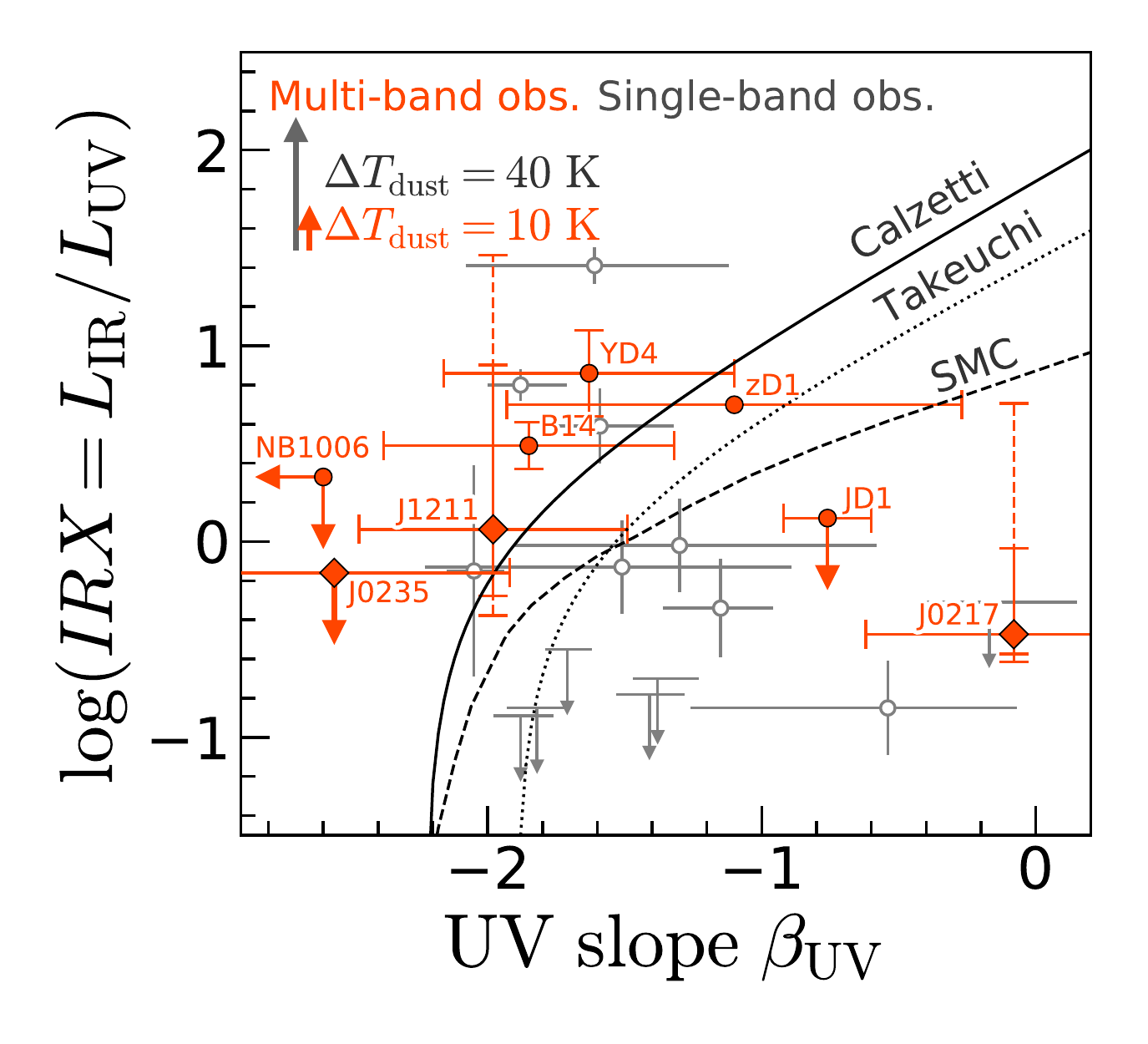}
 \end{center}
   \caption{
IRX as a function of the UV slope $\beta_\m{UV}$.
The red diamonds represent our targets at $z\sim6$, and the red circles show the IRX values of $z>7$ galaxies with multi-band observations taken from \citet{2019PASJ...71...71H}.
The dashed error bars show the range of IRX within the $1\sigma$ confidence regions in the error contours in Figure \ref{fig_fit_mbb}.
The gray circles and upper limits denote other $z>5$ galaxies without temperature determinations \citep{2017ApJ...845...41B}.
The solid, dotted, and dashed curves show the Calzetti extinction curve \citep{1999ApJ...521...64M,2000ApJ...533..682C}, the modified curve by \citet{2012ApJ...755..144T}, and the SMC extinction curve \citep{1998ApJ...508..539P}, respectively.
The black and red upward arrows show offsets of IRX with higher dust temperatures by 40 K and 10 K, respectively.
   \label{fig_IRX}}
\end{figure}

\begin{figure*}
\begin{center}
  \begin{minipage}{0.32\hsize}
 \begin{center}
  \includegraphics[clip,bb=13 8 275 312,width=0.99\hsize]{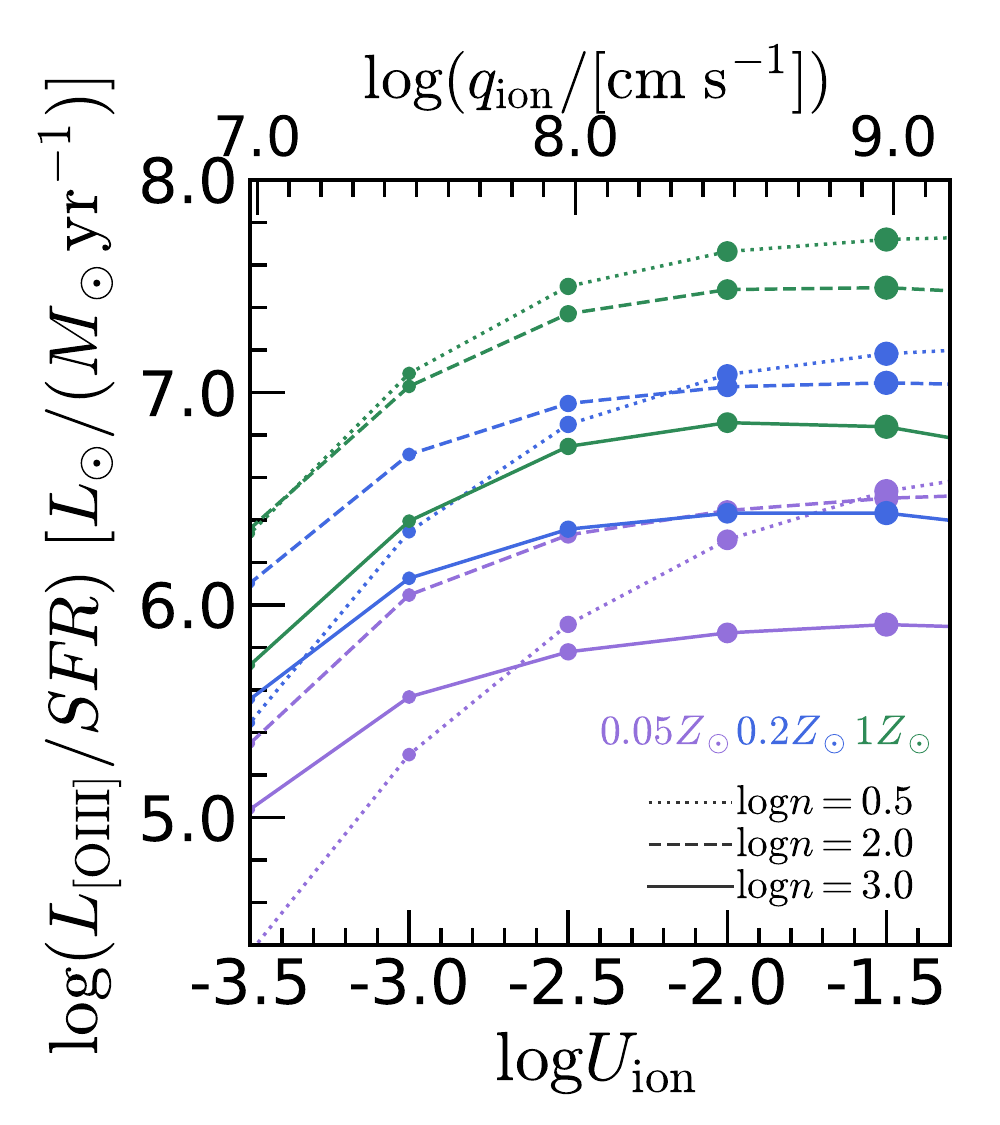}
 \end{center}
 \end{minipage}
 \begin{minipage}{0.32\hsize}
 \begin{center}
  \includegraphics[clip,bb=13 8 281 312,width=0.99\hsize]{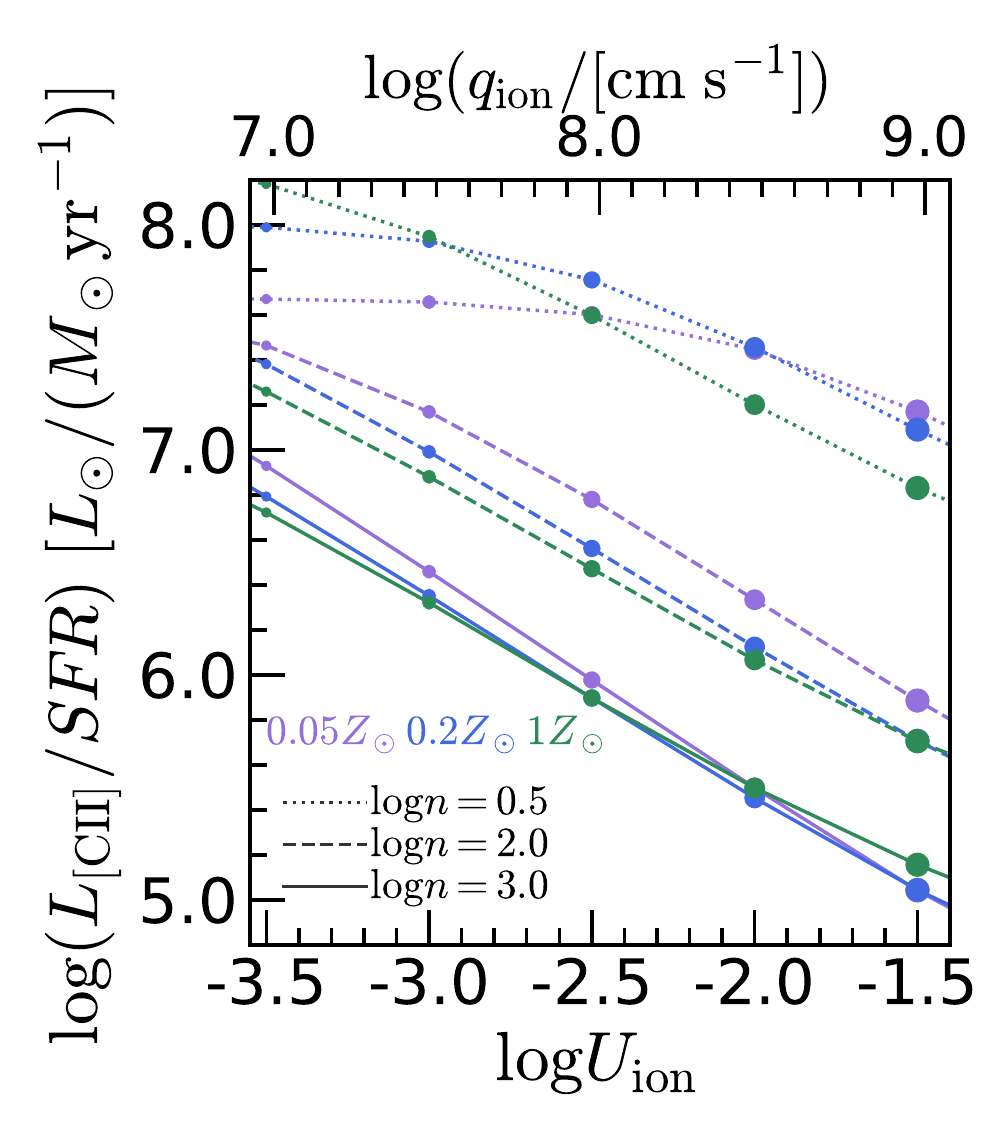}
 \end{center}
 \end{minipage}
 \begin{minipage}{0.32\hsize}
 \begin{center}
  \includegraphics[clip,bb=5 8 281 316,width=0.99\hsize]{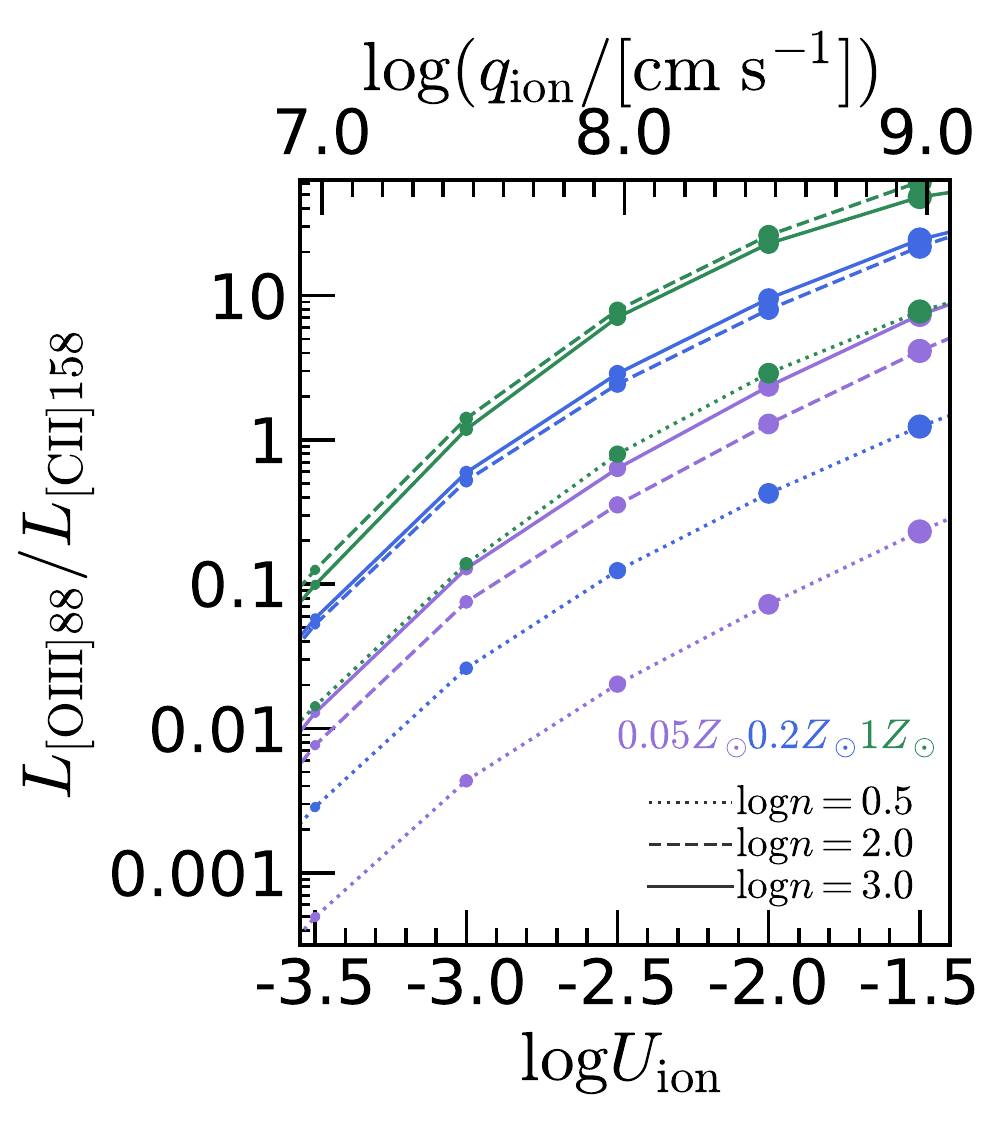}
 \end{center}
 \end{minipage}
 \end{center}
   \caption{
   Cloudy calculation results for $L_\m{[OIII]}/SFR$ ({\it left panel}), $L_\m{[CII]}/SFR$ ({\it middle panel}), and $L_\m{[OIII]}/L_\m{[CII]}$ ({\it right panel}) ratios as functions of the ionization parameter.
     The purple, blue, and green lines are results for metallicities of $Z=0.05\ Z_\odot$, $0.2\ Z_\odot$, and $1.0\ Z_\odot$, respectively.
  The dotted, dashed and solid lines correspond to densities of $\m{log}(n_\m{H}/[\m{cm^{-3}}])=0.5$, $2.0$, and $3.0$, respectively.
  The larger circles indicate higher ionization parameters, from $\m{log}U_\m{ion}=-4.0$ to $-0.5$ with a step size of 0.5.
   \label{fig_cl_OIII_CII_qion}}
\end{figure*}

Figure \ref{fig_IRX} shows the IRX values of our targets as a function of the UV slope $\beta_\m{UV}$.
We find that J1211-0118 follows the Calzetti IRX-$\beta_\m{UV}$ relation, similar to A1689-zD1 \citep{2015Natur.519..327W}, A2744-YD4 \citep{2017ApJ...837L..21L,2019MNRAS.487L..81L}, and B14-65666 \citep{2019PASJ...71...71H}.
The upper limit of J0235-0532 is consistent with both the Calzetti, Takeuchi, and SMC relations.
The IRX value of J0217-0208 is lower than the SMC relation, even if the IR luminosity is $L_\m{IR}=2.2\times10^{12}\ L_\odot$ (the dashed error bar).
Note that the wavelength ranges used to derive $\beta_\m{UV}$ differ from our sample and the others compared here (see also \citealt{2019PASJ...71...71H}); NIR imaging data are necessary to estimate $\beta_\m{UV}$ in the same wavelength ranges as the other studies.

There are several possibilities for the physical origins of the low IRX value of J0217-0208.
One is the geometry effect \cite[e.g.,][]{2017ApJ...847...21F}. 
Depending on the geometry and the viewing angle of the observer, the dust clouds might be spatially offset from the line-of-sight and the UV light can escape from the galaxy without being attenuated, which would cause the low IRX value.
AGN activity also can explain the offset in the IRX-$\beta_\m{UV}$ plot.
\citet{2018A&A...617A.131S} show that quasars have lower IRX values given the UV slope, similar to $z\sim5-6$ galaxies in \citet{2015Natur.522..455C}, because UV slopes of AGNs are redder than those of galaxies in $M_\m{UV}\gtrsim-24\ \m{mag}$. 
\citet{2017MNRAS.471.5018F} suggest that the high fraction of dust locked inside the molecular gas without internal illumination can explain the low IRX value.

\subsection{Dust Mass}
Based on the observed fluxes at $\sim160\ \mu\m{m}$ and the estimated temperatures, we estimate the dust masses in our targets.
Assuming a dust mass absorption coefficient $\kappa=\kappa_0(\nu/\nu_0)^{\beta_\m{d}}$, where $\kappa_0=10\ \m{cm^2\ g^{-1}}$ at 250 $\mu$m \citep{1983QJRAS..24..267H}, we estimate the dust masses to be $M_\m{dust}=3.0^{+10.5}_{-2.3}\times10^{7}\ M_\odot$, $<6.9\times10^{6}\ M_\odot\ (3\sigma)$, and $1.9^{+73.5}_{-1.6}\times10^{8}\ M_\odot$ for J1211-0118, J0235-0532, and J0217-0208, respectively.
Using an empirical relation between the UV magnitude and stellar mass for $z\sim6$ LBGs \citep{2016ApJ...825....5S}, the stellar masses of our targets are estimated to be $\sim3\times10^{10}\ M_\odot$.
The inferred dust-to-stellar mass ratios are $\m{log}(M_\m{dust}/M_*)\sim-3.0$, $<-3.6$, and $-2.2$ for J1211-0118, J0235-0532, and J0217-0208, respectively.
These dust-to-stellar mass ratios are relatively lower than the value for B14-65666 at $z=7$ \citep[$\m{log}(M_\m{dust}/M_*)\sim-1.9$;][]{2019PASJ...71...71H}, but within the range observed in $z\sim0$ galaxies \citep{2015A&A...582A.121R}.
These ratios can be reproduced by the supernova and AGB star model with grain growth in \citet{2015MNRAS.451L..70M}.
Note that our ALMA data probes 90, 120, and 160 $\mu$m in the rest-frame.
Since the longest wavelength corresponds to $\sim$20 K in the Wien law, comparable to the CMB temperature, no more cooler dust exists unless the CMB radiation was shielded.
\redcc{On the other hand, dust emission with the CMB temperature ($\sim20\ \m{K}$ at $z\sim6$) cannot be detected with interferometers like ALMA \citep{2013ApJ...766...13D}.
In the case that there is dust whose temperature is $\sim20\ \m{K}$ in the targeted galaxies, such dust would be missed in observations presented here, causing a large uncertainty in dust mass estimates.}

\section{Discussion}\label{ss_dis}

\subsection{Origin of High $L_\m{[OIII]}/L_\m{[CII]}$ Ratios at $z=6-9$}\label{ss_dis_OIIICII}

In Section \ref{ss_OIIICII}, we find that the {\sc [Oiii]/[Cii]} ratios of the $z=6-9$ galaxies are systematically higher than those of $z\sim0$ galaxies.
In order to discuss the origin of the high {\sc [Oiii]/[Cii]} ratios of the $z=6-9$ galaxies, we conduct model calculations including both H{\sc ii} regions and photodissociation regions (PDRs) using Cloudy \citep{1998PASP..110..761F,2017RMxAA..53..385F} version 17.01, following \citet{2011A&A...526A.149N,2012A&A...542L..34N}.
We include PDRs because {[C\sc ii]} emission mainly comes from PDRs \citep{2019A&A...626A..23C}.
We assume a pressure-equilibrium gas cloud with a plain-parallel geometry that is characterized by certain hydrogen gas densities at the ionization front ($n_\m{H}$), metallicity ($Z$), and ionization parameters $U_\m{ion}$.
Here we examine gas clouds with $\m{log}(n_\m{H}/[\m{cm^{-3}}])=0.5$, $2.0$, and $3.0$, and $Z=0.05$, $0.2$, and $1.0\ Z_\odot$ for $-4.0\leq\m{log}U_\m{ion}\leq-0.5$ with a step of 0.5.
Note that with the very high $U_\m{ion}$ value such as $\m{log}U_\m{ion}=-0.5$, the high pressure would squash the ionized gas \citep{2012ApJ...757..108Y}.
The input continuum is a spectrum of a burst star formation model with an age of 1 Myr and \citet{1955ApJ...121..161S} IMF with lower and upper mass cutoffs of $1\ M_\odot$ and $100\ M_\odot$ for binary populations, calculated with the BPASS v2.1 model \citep{2017PASA...34...58E}.
The stellar metallicity is equal to the gas metallicity.
The relative chemical composition of the gas cloud is scaled to the solar elemental abundance ratios except for helium, which follows an equation in \citet{2004ApJS..153...75G}.
The nitrogen abundance is also scaled to the solar abundance, although it is possible that the nitrogen abundance ratio changes due to its secondary element nature.
Orion-type graphite and silicate grains are included.
Calculations are stopped at the depth of $A_\m{V}=100\ \m{mag}$ to cover the whole {\sc[Cii]} emitting regions, following \citet{2005ApJS..161...65A}.

In the left panel of Figure \ref{fig_cl_OIII_CII_qion}, we plot $L_\m{[OIII]}/SFR$ ratios as a function of the ionization parameter.
The SFRs of the models are derived from H$\alpha$ luminosities using Equation (2) of \citet{1998ARA&A..36..189K}, and are converted to values in the \citet{2003PASP..115..763C} IMF by multiplying by $0.63$.
The uncertainty of the conversion in \citet{1998ARA&A..36..189K} is roughly $\sim30\%$ ($0.1\ \m{dex}$), which does not affect our conclusions.
The $L_\m{[OIII]}/SFR$ ratio increases with increasing ionization parameter, because more ionizing photons are used to produce $\m{O^{2+}}$.
Since the critical density of the {[\sc Oiii]}88$\mu$m is 510 cm$^{-3}$ for collision with electrons, the ratio decreases with increasing the density from $\m{log}(n_\m{H}/[\m{cm^{-3}}])=2.0$ to $\m{log}(n_\m{H}/[\m{cm^{-3}}])=3.0$.
Also, the ratio decreases with decreasing metallicity at $0.05<Z/Z_\odot<1.0$, because the lower abundance of the oxygen in the {\sc Hii} regions.

The middle panel of Figure \ref{fig_cl_OIII_CII_qion} shows $L_\m{[CII]}/SFR$ ratios as a function of the ionization parameter.
The $L_\m{[CII]}/SFR$ ratio decreases with increasing ionization parameter, because the volume of PDRs decreases due to the larger {\sc Hii} regions with more ionizing photons, and more $\m{C}^{+}$ are ionized to $\m{C}^{2+}$ \citep[see also][]{2019MNRAS.489....1F}.
We find that higher hydrogen densities also make lower $L_\m{[CII]}/SFR$ ratios, because the density in PDRs is higher than that in {\sc Hii} regions in the constant pressure assumption, and it exceeds the critical density of {[\sc Cii]}158$\mu$m for hydrogen, 2800 cm$^{-3}$.
The ratio does not strongly change with metallicity.
As discussed in \citet{2006ApJ...644..283K}, the carbon abundance is proportional to the metallicity, $Z$, while the PDR column density is proportional to $1/Z$, as long as shielding of FUV photons is dominated by dust (assuming a constant dust-to-metal ratio) and the gas column density is large enough.
Since the {\sc[Cii]} emission mainly comes from PDRs, the {\sc[Cii]} luminosity does not strongly depend on the metallicity ($L_\m{[CII]}\propto Z\times1/Z=\m{constant}$, see also \citealt{2019MNRAS.487.1689P,2019MNRAS.489....1F}).

We also show calculated {\sc [Oiii]/[Cii]} ratios as a function of the ionization parameter in the right panel of Figure \ref{fig_cl_OIII_CII_qion}.
The ratio increases with increasing ionization parameter or metallicity.
The ratio also increases with increasing the density from $\m{log}(n_\m{H}/[\m{cm^{-3}}])=0.5$ to $\m{log}(n_\m{H}/[\m{cm^{-3}}])=2.0$.

\begin{figure*}
 \begin{center}
  \includegraphics[clip,bb=24 2 674 319,width=0.98\hsize]{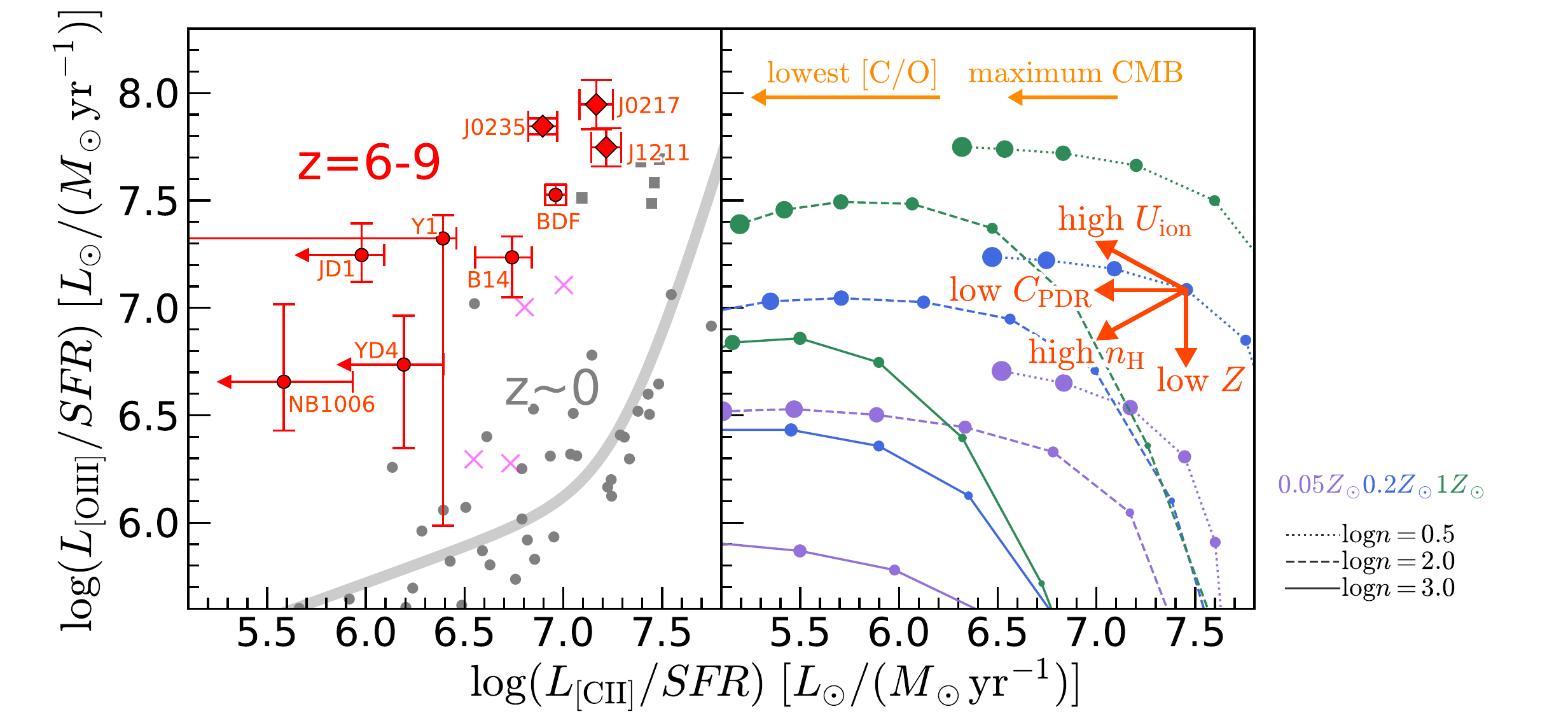}
 \end{center}
  \caption{
  {(\it Left panel}:) $L_\m{[OIII]}/SFR$ and $L_\m{[CII]}/SFR$ ratios.
  The red diamonds represent our targets at $z\sim6$, and the red circles are other $z=6-9$ LBGs and LAEs in the literature (see Table \ref{tab_lite}). 
  The gray squares and circles denote $z\sim0$ galaxies from the Dwarf Galaxy Survey \citep{2013PASP..125..600M,2014A&A...568A..62D,2015A&A...578A..53C} and GOALS \citep[][]{2010ApJ...715..572H,2017ApJ...846...32D}, respectively.
  The magenta crosses shows SMGs at $z=4.4$, $6.1$ and $6.9$ from \citet{2018Natur.560..613T,2019ApJ...876....1T}, \citet{2018ApJ...869L..22W}, and \citet{2018Natur.553...51M}, respectively (see Table \ref{tab_lite}).
  The $L_\m{[CII]}/SFR$ ratios of the $z=6-9$ galaxies are systematically lower than those of $z\sim0$ galaxies with similar $L_\m{[OIII]}/SFR$ ratios.
  ({\it Right panel}:)
  Cloudy calculation results for the $L_\m{[OIII]}/SFR$ and $L_\m{[CII]}/SFR$ ratios.
  The purple, blue, and green lines are results for metallicities of $Z=0.05\ Z_\odot$, $0.2\ Z_\odot$, and $1.0\ Z_\odot$, respectively.
  The dotted, dashed and solid lines correspond to densities of $\m{log}(n_\m{H}/[\m{cm^{-3}}])=0.5$, $2.0$, and $3.0$, respectively.
  The larger circles indicate higher ionization parameters, from $\m{log}U_\mathrm{ion}=-4.0$ to $-0.5$ with a step size of 0.5.
  The red arrows show directions of the shifts in the $L_\m{[OIII]}/SFR$-$L_\m{[CII]}/SFR$ plane by higher ionization parameter, lower PDR covering fraction, higher density, and lower metallicity (see text for details).
  The orange arrows indicate maximum shifts in $L_\m{[CII]}/SFR$ by the lower C/O ratio and the CMB attenuation effect.
   \label{fig_cl_OIIISFR_CIISFR}}
\end{figure*}

Since the SFR is an important parameter describing the amount of ionizing photons, we  need to compare both the line luminosities and SFRs of the observations with the models.
Thus we compare $L_\m{[OIII]}/SFR$ and $L_\m{[CII]}/SFR$ ratios of the $z=6-9$ galaxies with $z\sim0$ galaxies in the left panel of Figure \ref{fig_cl_OIIISFR_CIISFR}.
Our targets (J1211-0118, J0235-0532, and J0217-0208), MACS0416-Y1, B14-65666, and BDF-3299 are located in a region of $6.3<\m{log}(L_\m{[CII]}/SFR)<7.3$ and $7.2<\m{log}(L_\m{[OIII]}/SFR)<8.0$.
These galaxies (``{\sc [Cii]} detected galaxies'') have $\sim0.3-1\ \m{dex}$ lower $L_\m{[CII]}/SFR$ ratios than $z\sim0$ galaxies with similar $L_\m{[OIII]}/SFR$ ratios.
On the other hand, MACS1149-JD1, A2744-YD4, and SXDF-NB1006-2 are in $\m{log}(L_\m{[CII]}/SFR)<6.3$ and $6.6<\m{log}(L_\m{[OIII]}/SFR)<7.4$.
The $L_\m{[CII]}/SFR$ ratios of these galaxies (``{\sc [Cii]} undetected galaxies'') are $>1\ \m{dex}$ lower than $z\sim0$ galaxies with similar $L_\m{[OIII]}/SFR$ ratios.

\begin{figure*}
 \begin{center}
  \includegraphics[clip,bb=7 2 554 319,width=0.8\hsize]{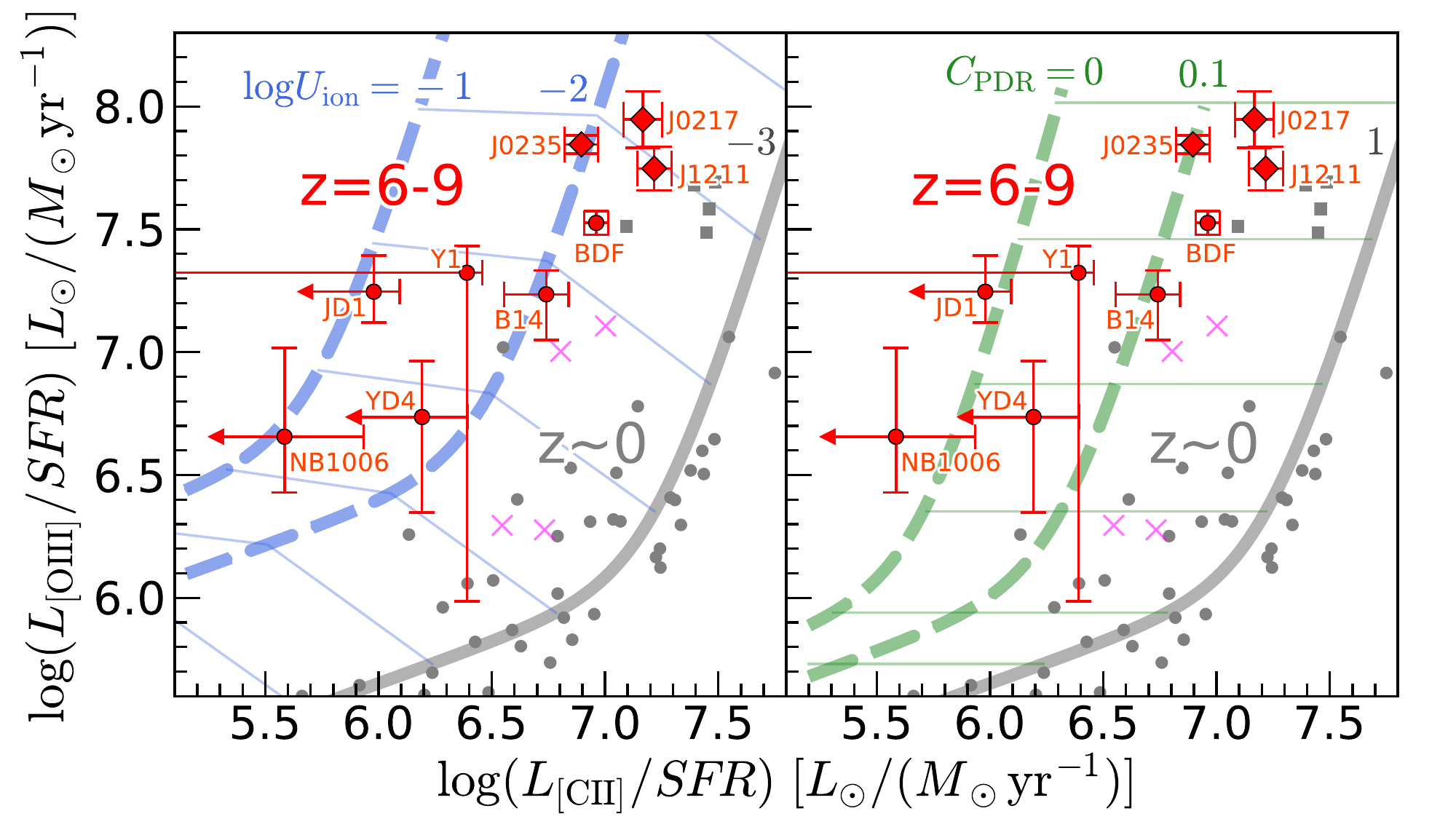}
 \end{center}
  \caption{
  Same as the left panel of Figure \ref{fig_cl_OIIISFR_CIISFR} but with the model curves with various ionization parameters (left) and PDR covering fractions (right).
  The solid gray curve shows the ratios for $\m{log}U_\m{ion}=-3$ and $C_\m{PDR}=1$.
  The dashed blue curves in the left panel are ratios in cases of $\m{log}U_\m{ion}=-2$ and $-1$ with $C_\m{PDR}=1$.
  The dashed green curves in the right panel indicate ratios in cases of $C_\m{PDR}=0.1$ and $0$ with $\m{log}U_\m{ion}=-3$.
  The high ionization parameter ($\times10-100$ higher $U_\m{ion}$ than $z\sim0$) or low PDR covering fraction ($C_\m{PDR}=0-0.1$) can reproduce the $z=6-9$ galaxies.
   \label{fig_cl_Uion_Cpdr}}
\end{figure*}

\begin{figure}
 \begin{center}
  \includegraphics[clip,bb=8 2 323 322,width=0.8\hsize]{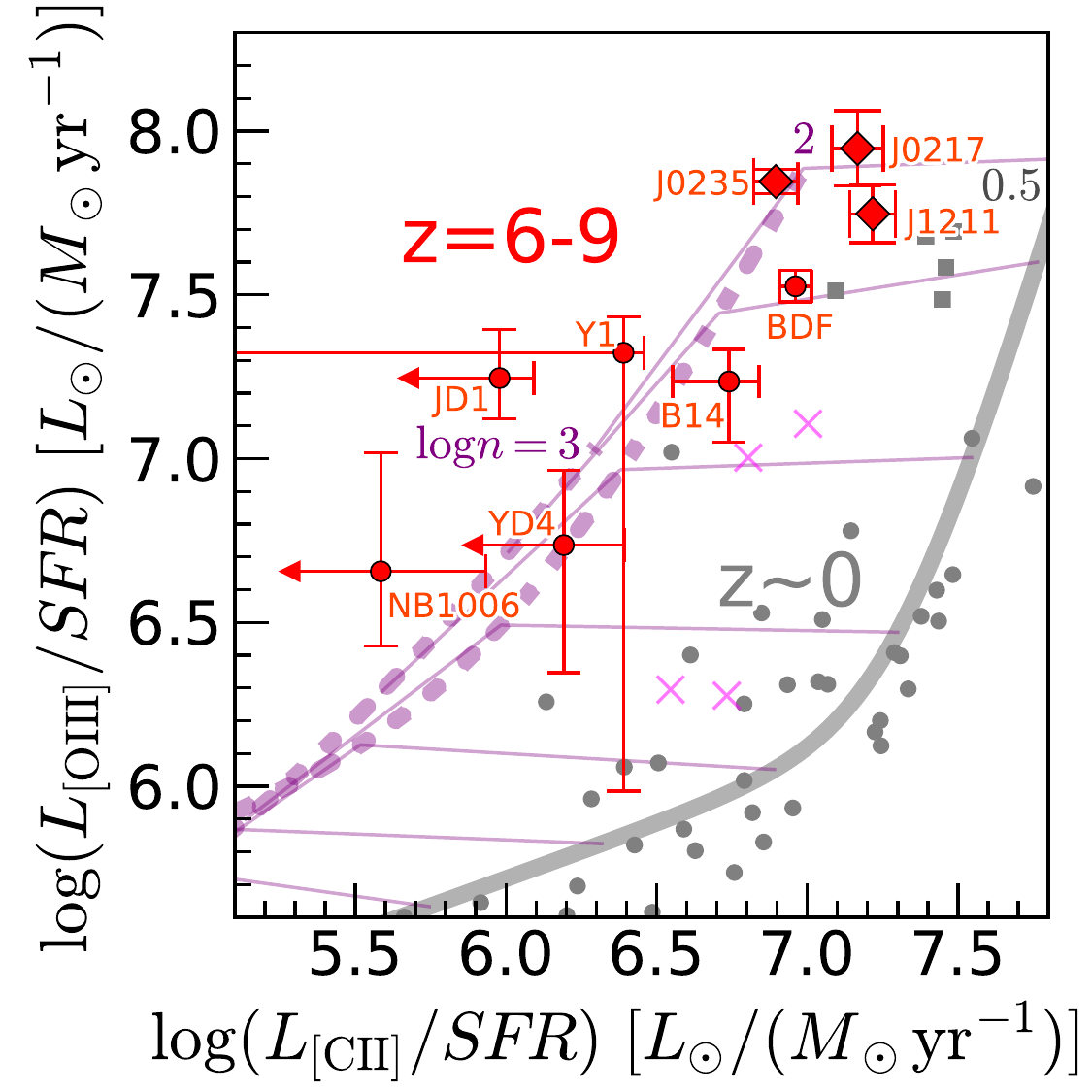}
 \end{center}
  \caption{
  Same as the Figure \ref{fig_cl_Uion_Cpdr} but with the model curves with various hydrogen densities.
  The solid gray curve shows the ratios for $\m{log}(n_\m{H}/[\m{cm^{-3}}])=0.5$.
  The dotted purple curves are ratios in cases of $\m{log}(n_\m{H}/[\m{cm^{-3}}])=2.0$ and $3.0$ with fixing other parameters.
  The higher densities ($\m{log}(n_\m{H}/[\m{cm^{-3}}])=2-3$) cannot explain some of the $z=6-9$ galaxies (e.g., MACS1149-JD1).
   \label{fig_cl_onlynH}}
\end{figure}

We plot results of the model calculations for $L_\m{[OIII]}/SFR$ and $L_\m{[CII]}/SFR$ ratios in the right panel of Figure \ref{fig_cl_OIIISFR_CIISFR}.
Based on these calculations, we discuss the following eight possibilities explaining the observational properties (i.e., $L_\m{[OIII]}/SFR$, $L_\m{[CII]}/SFR$, and high {\sc [Oiii]/[Cii]}) of the $z=6-9$ galaxies compared to the $z\sim0$ galaxies:
\begin{itemize}
\setlength{\itemsep}{0cm}
\setlength{\leftskip}{0.6cm}

\item[A)] Higher ionization parameter ($U_\m{ion}$)\\
Several studies suggest higher ionization parameters in higher redshift galaxies, compared to $z\sim0$ galaxies ($\m{log}U_\m{ion}\simeq-3.1$, e.g., \citealt{2014MNRAS.442..900N}).
As shown in Figures \ref{fig_cl_OIII_CII_qion} and \ref{fig_cl_OIIISFR_CIISFR}, with increasing $U_\m{ion}$, $L_\m{[OIII]}/SFR$ increases while $L_\m{[CII]}/SFR$ decreases.
Thus high ionization parameters can explain the $L_\m{[OIII]}/SFR$ and $L_\m{[CII]}/SFR$ ratios of both {\sc [Cii]} detected and undetected galaxies at $z=6-9$.
For example, as shown in the left panel of Figure \ref{fig_cl_Uion_Cpdr}, $2.0\ \m{dex}$ ($1.0\ \m{dex}$) higher ionization parameter can reproduce the $>1\ \m{dex}$ ($0.3-1\ \m{dex}$) systematic offsets of $L_\m{[CII]}/SFR$ seen in the {\sc [Cii]} undetected (detected) galaxies at $z=6-9$, compared to the $z\sim0$ galaxies, if we fix the metallicity and density.
Thus high ionization parameters would be an origin of the high {\sc [Oiii]/[Cii]} ratios of the $z=6-9$ galaxies.
The high ionization parameters in $z=6-9$ galaxies are possibly made by young (bursty) or low-metallicity stellar populations with more ionizing photons, or compact sizes of galaxies with high SFR surface densities \citep[e.g.,][]{2015ApJS..219...15S}, as predicted by theoretical simulations \citep[e.g.,][]{2017MNRAS.467.1300V,2018MNRAS.481L..84M}.
The high SFR surface density could ionize C$^+$ to C$^2+$ and make the {\sc Hii} regions overlapping each other like Figure \ref{fig_punch}b, increasing (decreasing) the volume of the {\sc Hii} regions (PDRs).
\redc{Observations for local starburst galaxies also report a positive correlation between the {\sc [Oiii]/[Cii]} ratio and the dust temperature \citep{2017ApJ...846...32D,2018ApJ...869L..22W}, possibly due to high ionization parameters in high {\sc [Oiii]/[Cii]} galaxies.}
Theoretical simulations of \citet{2019MNRAS.487.1689P} and \citet{2019MNRAS.489....1F} suggest that the burst of star formation enhance the ionization parameter and make the {\sc[Cii]} deficit, although it is not known whether the simulations also quantitatively reproduce the observed high {\sc [Oiii]/[Cii]} ratios.

\item[B)] Lower gas metallicity ($Z$)\\
Evolution of the mass-metallicity relation suggests lower metallicities in higher redshift galaxies \citep[e.g.,][]{2008A&A...488..463M}.
Figures \ref{fig_cl_OIII_CII_qion} and \ref{fig_cl_OIIISFR_CIISFR} indicate that with decreasing metallicity, only the $L_\m{[OIII]}/SFR$ ratio decreases, while the $L_\m{[CII]}/SFR$ ratio does not significantly change.
As discussed in the previous paragraph, the {\sc [Cii]} luminosity from the PDR does not strongly depend on the metallicity, as long as shielding of FUV photons is dominated by dust and the gas column density is large enough.
Thus changing the metallicity cannot reproduce the systematic offsets seen in $L_\m{[CII]}/SFR$ of the $z=6-9$ galaxies.

\item[C)] Higher density ($n_\m{H}$)\\
Several spectroscopic studies report high electron densities in high redshift galaxies \citep[e.g.,][]{2015MNRAS.451.1284S,2016ApJ...816...23S,2017ApJ...835...88K}.
As shown in Figures \ref{fig_cl_OIII_CII_qion} and \ref{fig_cl_OIIISFR_CIISFR}, both $L_\m{[OIII]}/SFR$ and $L_\m{[CII]}/SFR$ ratios decrease with increasing the hydrogen density due to the collisional de-excitation.
Since the $L_\m{[OIII]}/SFR$ ratios of the $z=6-9$ galaxies are comparable to $z\sim0$, higher densities cannot reproduce both $L_\m{[OIII]}/SFR$ and $L_\m{[CII]}/SFR$ of the $z=6-9$ galaxies simultaneously (Figure \ref{fig_cl_onlynH}).
More precisely, if we increase the density from $\m{log}(n_\m{H}/[\m{cm^{-3}}])=0.5$ to $2.0$, $L_\m{[OIII]}/SFR$ does not significantly change, while the $L_\m{[CII]}/SFR$ ratio decreases by $\sim1\ \m{dex}$, because the density in PDRs reaches the critical density of {\sc[Cii]}.
Thus as shown in Figure \ref{fig_cl_onlynH}, the increase of the density from $\m{log}(n_\m{H}/[\m{cm^{-3}}])=0.5$ to $2.0$ can reproduce a part of the $z=6-9$ galaxies, but cannot reproduce the $<1\ \m{dex}$ lower $L_\m{[CII]}/SFR$ ratios of some of the $z=6-9$ galaxies.
If we increase the density from $\m{log}(n_\m{H}/[\m{cm^{-3}}])=2.0$ to $3.0$, both the $L_\m{[CII]}/SFR$ and $L_\m{[OIII]}/SFR$ ratios decreases by $\sim1\ \m{dex}$.
As shown in Figure \ref{fig_cl_onlynH}, the increase of the density to $\m{log}(n_\m{H}/[\m{cm^{-3}}])=3.0$ can not reproduce some of the $L_\m{[OIII]}/SFR$ ratios of the $z=6-9$ galaxies (e.g., MACS1149-JD1).
In order to explain the properties of $z=6-9$ galaxies with only the increase of the density alone, we need much higher density in PDRs with relatively lower density in {\sc Hii} regions.
However, theoretical simulations \citep[e.g.,][]{2005ApJ...623..917H,2010ApJ...716.1191W} predict weaker contrat of the density between {\sc Hii} regions and PDRs than the constant pressure assumption used in our calculations \citep[see Figure 2 in][]{2019A&A...626A..23C}.

\item[D)] Lower $\m{C/O}$ ratio\\
Spectroscopic studies suggest that the C/O abundance ratios of high redshift galaxies are lower than the solar abundance ratio \citep[e.g.,][]{2016ApJ...826..159S}.
Since {\sc [Cii]} and {\sc [Oiii]} luminosities depend on the carbon and oxygen abundances, respectively, lower C/O ratios of $z=6-9$ galaxies may explain the lower $L_\m{[CII]}/SFR$ ratio.
\citet{2016ApJ...826..159S} report that C/O abundance ratio of $z\sim2$ galaxies is $\sim50\%$ of the solar abundance ratio.
\citet{2017MNRAS.464..469S} also suggest a $\sim50\%$ solar abundance ratio for a $z=7.7$ galaxy.
The low C/O ratio would be due to young stellar population in high redshift galaxies.
Oxygen is mainly produced by core-collapse supernovae (SNe), and therefore has the shortest formation timescales \citep{2019A&ARv..27....3M}.
On the other hand, carbon has contributions from both core-collapse and type-Ia SNe and from AGB stars, and its average formation timescale is longer than that of oxygen.
Thus the low C/O ratio indicates young stellar age with the ongoing production of carbon.
The lowest C/O would be $\m{C/O}\simeq0.1(\m{C/O})_\odot$ based on theoretical calculations \citep[e.g.,][]{2019A&ARv..27....3M} and observations \citep[e.g.,][]{2016ApJ...832..171T,2017MNRAS.467..802C}.
We conduct model calculations again with 50\% and 10\% solar abundance ratio, $[\m{C/O}]=-0.3$ and $-1.0$, with fixing the oxygen abundance.
We find that the $L_\m{[OIII]}/SFR$ ratio does not change, while the $L_\m{[CII]}/SFR$ ratio decreases by 0.3 dex and 0.9 dex, for $[\m{C/O}]=-0.3$ and $-1.0$, respectively.
$[\m{C/O}]=-0.3$ would explain the 0.3 dex lower $L_\m{[CII]}/SFR$ ratio for a part of the {\sc [Cii]} detected galaxies at $z=6-9$.
For the {\sc [Cii]} undetected galaxies, $[\m{C/O}]=-1.0$ cannot reproduce the $>1\ \m{dex}$ lower $L_\m{[CII]}/SFR$ ratio.
Therefore, low C/O ratios would reproduce a part of the $z=6-9$ galaxies, but not all of them.
Note that theoretical simulations in \citet{2020arXiv200101853A} also indicate that low C/O ratios partly contribute to the observed high {\sc [Oiii]/[Cii]} ratios.

\begin{figure*}
 \begin{center}
  \includegraphics[clip,bb=94 114 783 405,width=0.8\hsize]{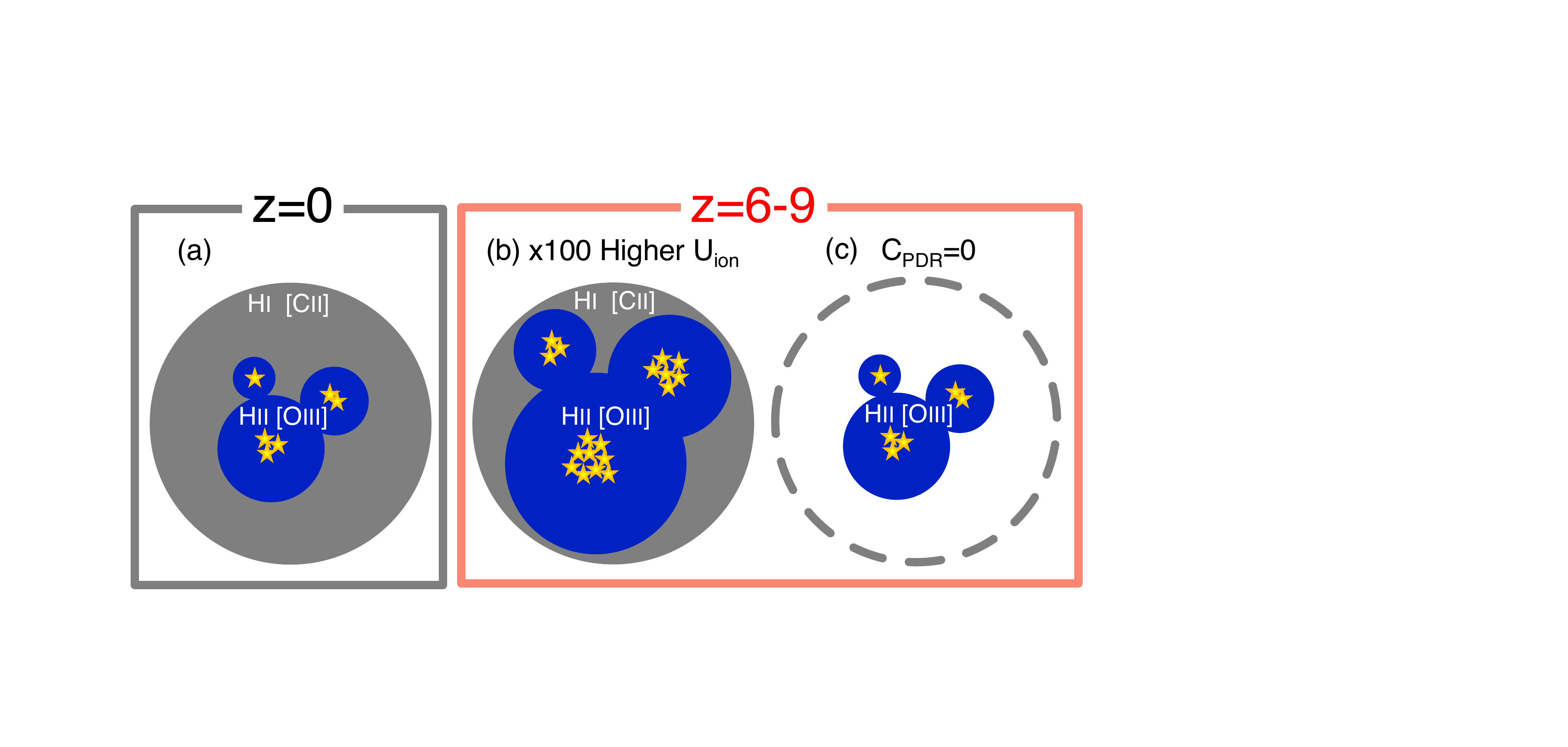}
 \end{center}
  \caption{
Schematic illustrations of the {\sc Hii} regions and PDRs.
The {\sc Hii} regions including the {\sc Oiii} emitting gas are presented with blue, and the outer PDRs are shown with grey.
Yellow stars are central ionizing sources.
(a) A case of an ionization bounded nebula whose radius is determined by the ionization equilibrium for $z\sim0$ galaxies.
(b) A case of a high ionization parameter ($U_\m{ion}$) for $z\sim6-9$ galaxies.
The young stellar population or the compact size would make high $U_\m{ion}$ and larger {\sc Hii} regions relative to PDRs, resulting in high {\sc[Oiii]}/{\sc[Cii]} ratios.
(c) A case of a low PDR covering fraction ($C_\m{PDR}$) for $z\sim6-9$ galaxies.
This illustrates an extreme case of $C_\m{PDR}=0$.
A low PDR covering fraction makes a low {\sc[Cii]} luminosity.
Note that in real galaxies there would be some clumpy molecular gas in the {\sc Hii} regions due to the self shielding like the Orion nebula, leading to next star formation.
   \label{fig_punch}}
\end{figure*}

\begin{figure}
 \begin{center}
  \includegraphics[clip,bb=739 115 1018 438,width=0.6\hsize]{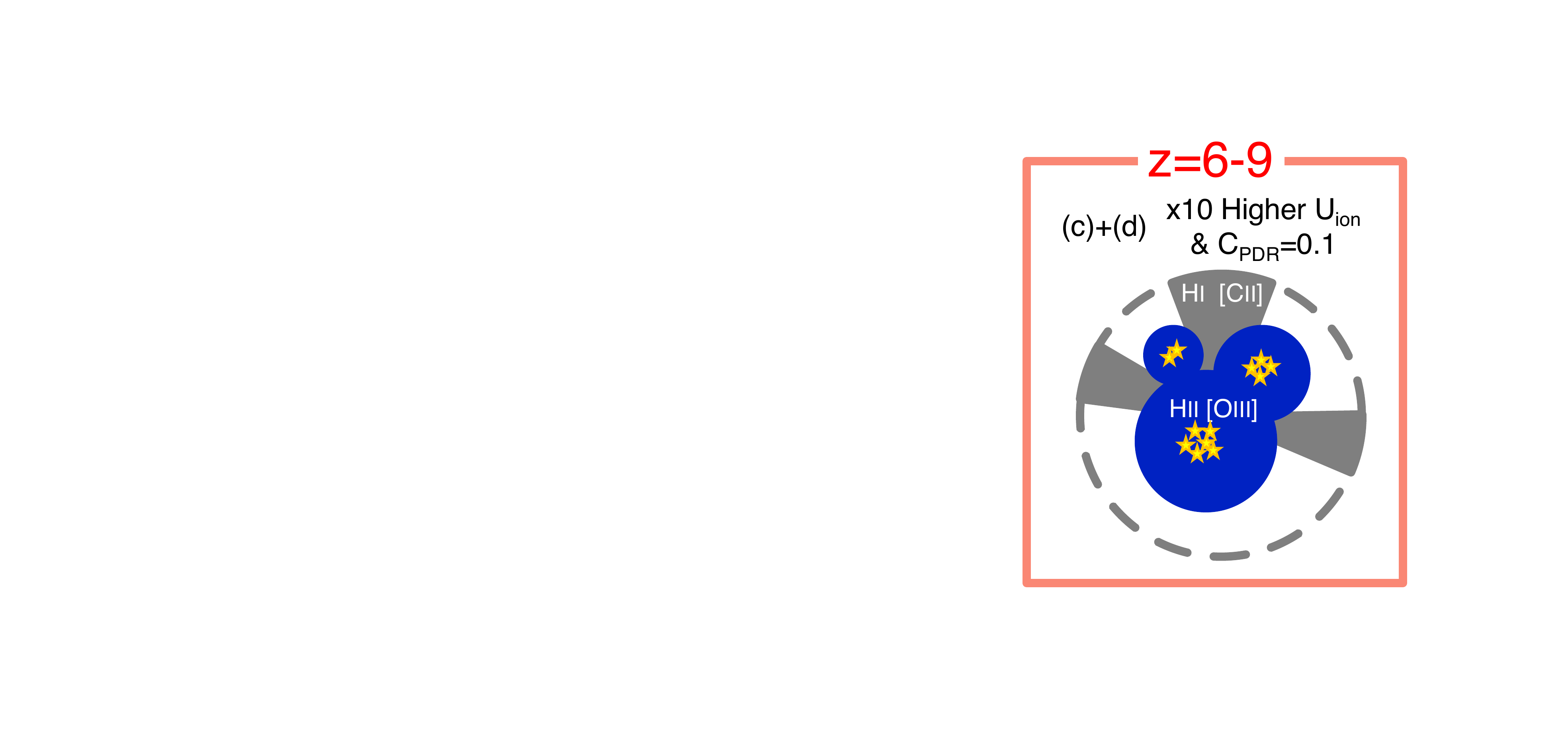}
 \end{center}
  \caption{
Same as Figure \ref{fig_punch} but for a case of the combination of case (b) and (c); $\times10$ higher $U_\m{ion}$ and $C_\m{PDR}=0.1$.
   \label{fig_punch_combination}}
\end{figure}

\item[E)] Lower PDR covering fraction ($C_\m{PDR}$)\\
{\sc [Oiii]} emission comes from {\sc Hii} regions, while {\sc [Cii]} emission mainly comes from PDRs \citep[e.g.,][]{2006ApJ...644..283K,2019A&A...626A..23C}.
Thus the covering fraction of the PDR (and the neutral H{\sc i} gas) with respect to the H{\sc ii} region, $C_\m{PDR}$, is an important parameter to determine the {\sc[Cii]} luminosity and the {\sc [Oiii]/[Cii]} ratio.
Following \citet{2019A&A...626A..23C}, we define the PDR covering fraction, $C_\m{PDR}$, as a fraction of the number of sightlines with PDRs to that with {\sc Hii} regions,
\begin{equation}
C_\m{PDR}=\frac{\m{No.\ of\ sightlines\ with\ PDRs}}{\m{No.\ of\ sightlines\ with\ {H\textsc{ii}}\ regions}}.
\end{equation}
$C_\m{PDR}$ is a parameter varying from zero to unity that corresponds to a linear scaling of the PDR intensities.
If $C_\m{PDR}=1$, all of the {\sc Hii} regions are covered with PDRs (cases (a) and (b) in Figure \ref{fig_punch}).
If $C_\m{PDR}=0$, none of the {\sc Hii} regions are covered with PDRs (density-bounded, case (c) in Figure \ref{fig_punch}).
For galaxies with $Z\sim0.05Z_\odot$, the fraction of {\sc [Cii]} emission that comes from {\sc Hii} regions is $\sim1\%$ \citep{2019A&A...626A..23C}.
As shown in the right panel of Figure \ref{fig_cl_Uion_Cpdr}, if we consider an extreme case of $C_\m{PDR}=0$, the {\sc[Cii]} luminosity decreases by $\sim99\%$ ($-2\ \m{dex}$), which can explain the systemically lower $L_\m{[CII]}/SFR$ ratios of the $z=6-9$ galaxies compared to the $z\sim0$ galaxies.
Thus the lower PDR covering fraction could be an origin of the high {\sc [Oiii]/[Cii]} ratio of high redshift galaxies.
The low PDR covering fraction may be due to compact sizes of galaxies or outflow, consistent with observational evidence of higher outflow velocity in higher redshift galaxies \citep{2017ApJ...850...51S,2019arXiv190403106S}.
These low $C_\m{PDR}$ galaxies may allow a significant escape of Lyman continuum photons like a density-bounded nebula \citep[Figure \ref{fig_punch} (c);][]{2014MNRAS.442..900N}, and can be important contributors for the cosmic reionization.

\item[F)] CMB attenuation effect\\
At $z=6-9$, the CMB radiation affects FIR emission lines from galaxies, because the CMB temperature is around $20-30\ \m{K}$, sometimes comparable to excitation temperatures of the emission lines  \citep[e.g.,][]{2013ApJ...766...13D,2014ApJ...784...99G,2015ApJ...813...36V,2015MNRAS.453.1898P,2017MNRAS.465.2540P,2018A&A...609A.130L}.
The CMB effect does not have a significant impact on emission from {\sc Hii} regions (e.g., {\sc [Oiii]}) due to its high excitation temperature, but becomes important for emission from PDRs, such as {\sc [Cii]} in diffuse PDRs.
The high temperature of the CMB can heat the $\m{C^+}$ gas, while the CMB represents a strong background.
As a result the {\sc[Cii]} emission appears to be weaker, so called the CMB attenuation.
Recent calculations by \citet{2018A&A...609A.130L} show that {\sc [Cii]} emission is significantly attenuated by the CMB \citep[see also][]{2014ApJ...784...99G}.
However, as discussed in \citet{2019MNRAS.487L..81L}, the maximum effect of the CMB attenuation is $\sim0.5\ \m{dex}$.
Thus the CMB attenuation effect explains the 0.3-0.5 dex lower $L_\m{[CII]}/SFR$ ratio of a part of the {\sc [Cii]} detected galaxies, but cannot explain the $>1.0$ dex lower ratio of the {\sc [Cii]} undetected galaxies at $z=6-9$.

\item[G)] Spatially extended {\sc[Cii]} halo\\
\citet{2019arXiv190206760F} detect spatially extended {\sc[Cii]} halos around galaxies at $z=5-7$.
Such {\sc[Cii]} halos could be missed in the flux measurements, resulting in the lower $L_\m{[CII]}/SFR$ ratios of the $z=6-9$ galaxies.
However, we use the $2\arcsec$-radius apertures for the flux measurements, which cover the total flux of the {\sc[Cii]} halo.
In addition, this effect is not enough to explain the low $L_\m{[CII]}/SFR$ ratios in other studies.
For example, if the {\sc[Cii]} emission flux is measured in a $\sim0.\carcsec7$-diameter aperture, comparable to beam sizes in \citet{2016Sci...352.1559I} and \citet{2019MNRAS.487L..81L}, the total flux is underestimated only up to $\sim0.6\ \m{dex}$, assuming the radial profile of the {\sc[Cii]} halo in \citet{2019arXiv190206760F}.
Thus the extended {\sc[Cii]} halo cannot explain the low $L_\m{[CII]}/SFR$ ratios of the $z=6-9$ galaxies.

\item[H)] Inclination effect\\
\citet{2019MNRAS.487.3007K} suggest that inclination effects are responsible for some of the non-detections of {\sc[Cii]} emission in $z>6$ galaxies.
Mock ALMA simulations in \citet{2019MNRAS.487.3007K} show that {\sc[Cii]} is detected at $>5\sigma$ when seen face-on, while in the edge-on case it remains undetected because the larger intrinsic FWHM ($\sim600\ \m{km\ s^{-1}}$) pushes the line peak flux below the detection limit.
However, in this case, the {\sc[Oiii]} emission line is also difficult to be detected, resulting no significant change in the {\sc[Oiii]}/{\sc[Cii]} ratio.
In addition, the observed FWHMs of the emission lines are $200-400\ \m{km\ s^{-1}}$, not as large as $600\ \m{km\ s^{-1}}$ predicted in \citet{2019MNRAS.487.3007K}.
Thus the inclination effect cannot explain the high {\sc[Oiii]}/{\sc[Cii]} ratio of $z=6-9$ galaxies.

\end{itemize}

Based on these discussions, we conclude that A) higher ionization parameter or E) lower PDR covering fraction can explain the properties of the $z=6-9$ galaxies including the high {\sc [Oiii]/[Cii]} ratios and low $L_\m{[CII]}/SFR$ ratios.
Figure \ref{fig_punch} illustrates these cases.
Compared to the $z\sim0$ galaxies (case (a)), we need the $\times100$ higher ionization parameter (case (b)), or very low PDR covering fraction ($C_\m{PDR}=0$) like a density-bounded nebula (case (c)).
The middle of these two cases, i.e., $\times\sim10$ higher $U_\m{ion}$ and $C_\m{PDR}\sim0.1$, is also possible (Figure \ref{fig_punch_combination}).
We find that C) higher density, D) lower $\m{C/O}$ ratio, and F) CMB attenuation effect can reproduce only a part of the $z=6-9$ galaxies, but these effects cannot explain all of the $z=6-9$ galaxies with {\sc [Oiii]} and {\sc [Cii]} observations by itself.
The combination of two of these effects can reproduce the observed properties of the $z=6-9$ galaxies.

\subsection{{\sc [Oiii]/[Nii]} Ratios and Ionization parameters}
\begin{figure}
\begin{center}
  \includegraphics[clip,bb=5 8 281 316,width=0.8\hsize]{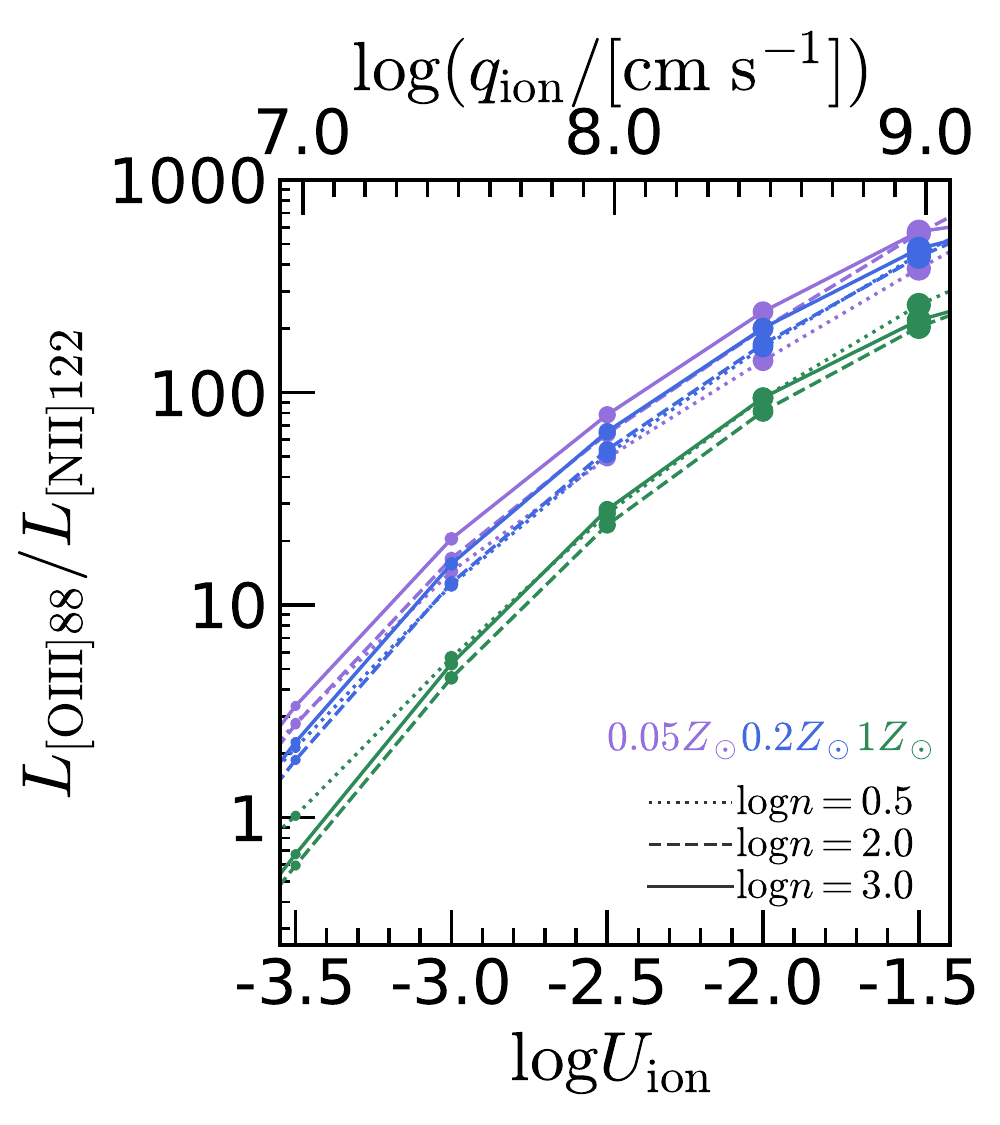}
 \end{center}
\caption{
   Cloudy calculation results for $L_\m{[OIII]}/L_\m{[NII]}$ ratios as a function of the ionization parameter.
     The purple, blue, and green lines are results for metallicities of $Z=0.05\ Z_\odot$, $0.2\ Z_\odot$, and $1.0\ Z_\odot$, respectively.
  The dotted, dashed and solid lines correspond to densities of $\m{log}(n_\m{H}/[\m{cm^{-3}}])=0.5$, $2.0$, and $3.0$, respectively.
  The larger circles indicate higher ionization parameters, from $\m{log}U_\mathrm{ion}=-4.0$ to $-0.5$ with a step size of 0.5.
    The solar N/O abundance ratio is assumed.
   \label{fig_cl_OIINII_qion}}
\end{figure}

Figure \ref{fig_cl_OIINII_qion} shows the {\sc[Oiii]}88$\mu$m/{\sc[Nii]}122$\mu$m ratio as a function of the ionization parameter from our Cloudy calculations.
Since the critical densities of {\sc[Nii]}122$\mu$m and {\sc[Oiii]}88$\mu$m are similar (310 and 510 cm$^{-3}$, respectively), the {\sc[Oiii]}/{\sc[Nii]} ratio is not sensitive to the electron density, but
sensitive to the ionization parameter, due to the different ionization potential between $\m{O^{2+}}$ and $\m{N^+}$.
Figure \ref{fig_cl_OIINII_qion} also indicates that the {\sc[Oiii]}/{\sc[Nii]} ratio becomes lower with higher metallicity due to less high energy photons ionizing $\m{O}^+$ to $\m{O}^{2+}$.
The N/O abundance ratio is assumed to be the solar abundance ratio.
This assumption is reasonable as long as we are focusing on galaxies with $Z<0.23\ Z_\odot$ \citep{2002ApJS..142...35K}.
Here we discuss whether the non-detections of {\sc [Nii]} in our targets are consistent with high $U_\m{ion}$ or low $C_\m{PDR}$ scenarios suggested by the high {\sc [Oiii]}/{\sc [Cii]} ratio in Section \ref{ss_dis_OIIICII}.
The observed {\sc[Oiii]}/{\sc[Nii]} ratios are $>5.8$, $>3.2$, and $>13.8$ for J1211-0118, J0235-0532, and J0217-0208, respectively, corresponding to ionization parameters of $\m{log}U_\m{ion}>-3.1$.
This is consistent with the high $U_\m{ion}$ scenario ($\m{log}U_\m{ion}>-3$), and the low $C_\m{PDR}$ scenario with fixed $U_\m{ion}=-3$.
\redc{Note that the {\sc[Oiii]}/{\sc[Nii]} ratio also depends on the excitation temperature, because the {\sc[Oiii]}88$\mu$m line is the lower transition ($^3P_1\rightarrow^3P_0$) while the {\sc[Nii]}122$\mu$m line is the upper transition ($^3P_2\rightarrow^3P_1$).
Nonetheless, recent ALMA observations report weak {\sc[Nii]}205$\mu$m ($^3P_1\rightarrow^3P_0$) lines in LBGs at $z\sim5$ probably due to high ionization parameters \citep{2016ApJ...832..151P,2019ApJ...882..168P}, consistent with our results.}

\begin{figure*}
\begin{center}
   \begin{minipage}{0.55\hsize}
 \begin{center}
  \includegraphics[clip,bb=16 19 325 287,width=0.85\hsize]{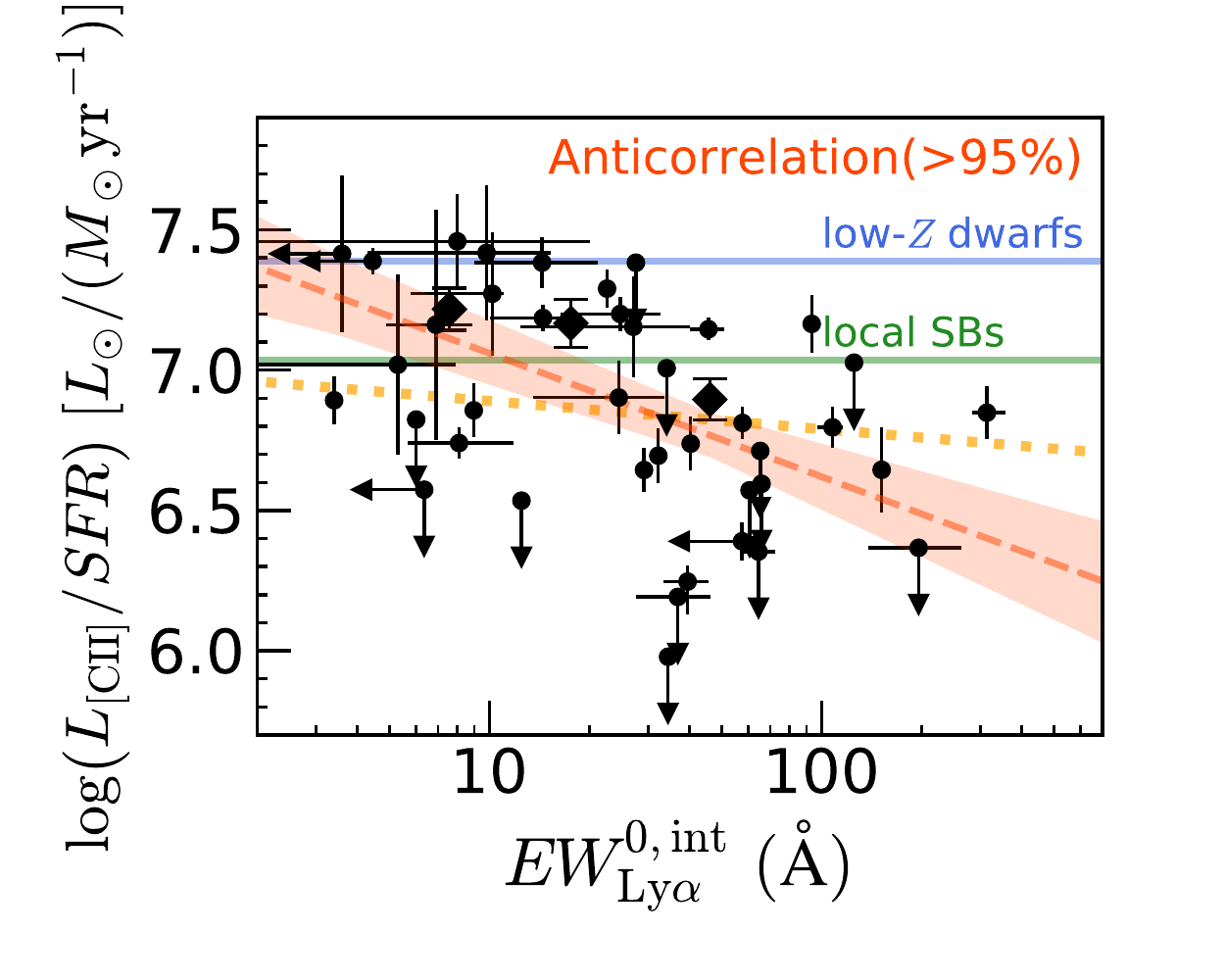}
 \end{center}
 \end{minipage}
 \begin{minipage}{0.43\hsize}
 \begin{center}
  \includegraphics[clip,bb=7 8 287 284,width=0.85\hsize]{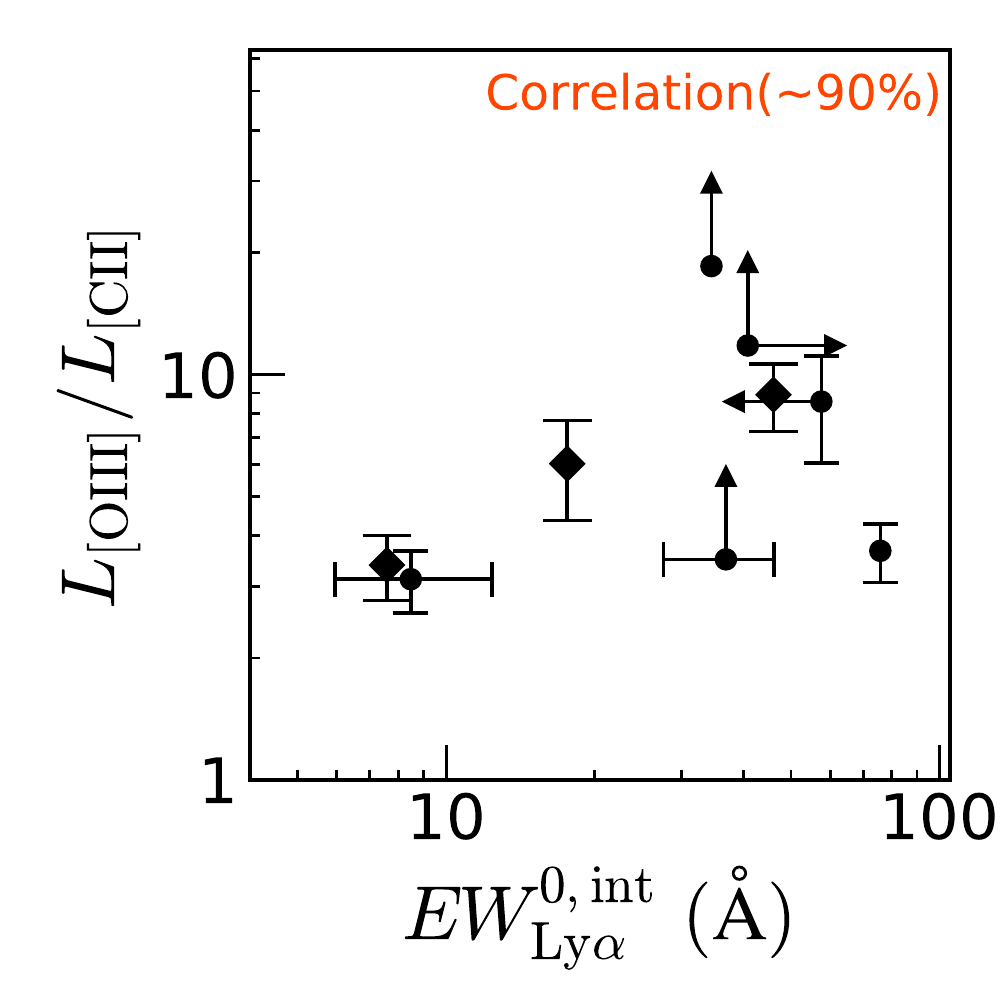}
 \end{center}
 \end{minipage}
 \end{center}
   \caption{
   ({\it Left panel:}) 
   $L_\m{[CII]}/SFR$ as a function of the rest-frame Ly$\alpha$ EW. 
The black diamonds represent our targets at $z\sim6$, and the black circles are other $z=5-9$ galaxies in the literature (see Table \ref{tab_lite}).
 We find an anticorrelation between $L_\m{[CII]}/SFR$ and the rest-frame intrinsic Ly$\alpha$ EW, $EW^{0,\m{int}}_\m{Ly\alpha}$, at the $>95\%$ significance level.
 The red dashed line and the shaded regions are the best-fit relation and its uncertainty.
The blue and green horizontal lines show the $L_\m{[CII]}/SFR$ ratios for low-metallicity dwarf galaxies and local starburst galaxies in \citet{2014A&A...568A..62D}, respectively, for $SFR=10\ M_\odot\ \m{yr^{-1}}$.
\redc{The orange dotted line is a fitting function in \citet{2020arXiv200200979S}.}
The rest-frame intrinsic Ly$\alpha$ EW, $EW^{0,\m{int}}_\m{Ly\alpha}$ is calculated from the observed rest-frame Ly$\alpha$ EW following Equations (13)-(17) in \citet{2018ApJ...859...84H}.
  ({\it Right panel:}) 
  {\sc [Oiii]}/{\sc [Cii]} ratios as a function of the rest-frame Ly$\alpha$ EW.
  The black diamonds represent our targets at $z\sim6$, and the black circles are other $z=6-9$ galaxies in the literature (see Table \ref{tab_lite}).
  We identify a correlation between $L_\m{[OIII]}/L_\m{[CII]}$ and $EW^{0,\m{int}}_\m{Ly\alpha}$ at the
$\sim90\%$ significance level.
   \label{fig_CIISFR_EW}}
\end{figure*}

\subsection{Correlations with Ly$\alpha$ and Physical Interpretations}\label{ss_dis_CIISFR}

\citet{2018ApJ...859...84H} report the anti-correlation between the $L_\m{[CII]}/SFR$ ratio and the Ly$\alpha$ EW, indicating the {\sc[Cii]} deficit in LAEs.
Our targets show a good anti-correlation with $(L_\m{[CII]}/SFR, EW^0_\m{Ly\alpha})=(1.6\times10^7,6.9)$, $(1.5\times10^7,15)$, and $(0.8\times10^7,41)$ in units of $(L_\odot/(M_\odot\ \m{yr}^{-1}), \m{\AA})$ for J1211-0118, J0217-0208, and J0235-0532, respectively.
Including the results of our targets and the literature, we find that the anti-correlation holds with the $99.6\%$ significance level with the Kendall’s tau test (see the left panel of Figure \ref{fig_CIISFR_EW}).
In the left panel of Figure \ref{fig_CIISFR_EW}, we also plot the following power-law function:
\begin{equation}\label{eq_CIISFR_EW}
\m{log}(L_\m{[CII]}/SFR)=-0.44\times\m{log}EW^\m{0,int}_\m{Ly\alpha}+7.5.
\end{equation}
\redc{In order to investigate whether this anti-correlation is made by a selection effect in which high $EW^0_\m{Ly\alpha}$ galaxies tend to be less luminous (less massive) galaxies, we calculate median SFRs of galaxies in $EW^\m{0,int}_\m{Ly\alpha}$ bins.
The calculated median SFRs are $21$, $26$, and $18\ M_\odot\ \m{yr^{-1}}$ for $EW^\m{0,int}_\m{Ly\alpha}<10\ \m{\AA}$, $10\ \m{\AA}<EW^\m{0,int}_\m{Ly\alpha}<50\ \m{\AA}$, and $50\ \m{\AA}<EW^\m{0,int}_\m{Ly\alpha}$, respectively.
Because the SFR values are not significantly different, the anti-correlation would not be explained by the selection effect.}

\redc{Recently \citet{2020arXiv200200979S} studied the $L_\m{[CII]}/SFR-EW^0_\m{Ly\alpha}$ relation using the ALPINE sample, and reported a weaker dependence of $L_\m{[CII]}/SFR$ on $EW^0_\m{Ly\alpha}$ with a slope of $-0.11$ in Equation (\ref{eq_CIISFR_EW}) compared to our result.
They discussed that the difference between their result and our result can be explained by the sample difference, their use of a single IMF, a consistent use of calibrated SFRs (no SED-based SFRs) following \citet{2019ApJ...881..124M}, and/or an adoption of conservative line widths to derive $L_\m{[CII]}$ upper limits.
Regarding the IMF, all of the SFRs used in this study are based on the \citet{2003PASP..115..763C} IMF and corrected for the IMF differences.
To check whether the difference in the $L_\m{[CII]}/SFR-EW^0_\m{Ly\alpha}$ relations is due to the methods of estimating SFRs or deriving $L_\m{[CII]}$ upper limits, we calculate the significance level of the anti-correlation from our sample in two cases of using SFRs in \citet{2019ApJ...881..124M}, and excluding upper limit data.
We find that the anti-correlation is significant with the $99.5\%$ and $99.4\%$ significance levels (with slopes of $-0.50$ and $-0.36$ in Equation (\ref{eq_CIISFR_EW})) in these two cases, respectively, indicating that the difference of the method cannot explain the difference in the $L_\m{[CII]}/SFR-EW^0_\m{Ly\alpha}$ relations.
We also calculate the significance using our sample without correction for IGM absorption in \citet{2018ApJ...859...84H}, and find that the anti-correlation still holds at the $95\%$ level (with a slope of $-0.28$).
These results indicate that the difference of the $L_\m{[CII]}/SFR-EW^0_\m{Ly\alpha}$ relation in this study and \citet{2020arXiv200200979S} is due to the sample difference.
One possible explanation is a difference of redshifts in the samples.
Our galaxies are mainly at $z>6$, while the ALPINE sample in \citet{2020arXiv200200979S} is based on galaxies at $z=4-6$.
The other possible explanation is a difference of specific SFRs.
Our galaxy sample contains all of the results reported, while the ALPINE sample is selected to be located on the star-formation main sequence, which could miss galaxies with high specific SFRs such as LAEs \citep{2010MNRAS.402.1580O,2016ApJ...817...79H,2018ApJ...859...84H}.
}

The right panel of Figure \ref{fig_CIISFR_EW} shows the {\sc [Oiii]/[Cii]} ratios as a function of the Ly$\alpha$ EW.
We find that galaxies with higher $EW^\m{0,int}_\m{Ly\alpha}$ tend to have higher {\sc [Oiii]/[Cii]} ratios at the $\sim90\%$ significance level \citep[see also][]{2019PASJ...71...71H}.
As discussed in Section \ref{ss_dis_OIIICII}, high {\sc[Oiii]/[Cii]} ratios would be due to high ionization parameters or low PDR covering fraction.
In the high ionization parameter case, the intense radiation field would ionize $\m{C^{+}}$ and neutral hydrogen in PDRs, decreasing the {\sc [Cii]} emissivity and {\sc Hi} column density, and increasing the transmission of Ly$\alpha$.
In the low PDR covering fraction case, Ly$\alpha$ photons directly escape from {\sc Hii} regions that are not covered by PDRs.
Thus, the origin of the {\sc[Cii]} deficit in LAEs would be the high ionization parameter or low PDR covering fraction.

The low PDR covering fraction would also enhance the Lyman continuum photon escape from galaxies at the reionization epoch.
Indeed, recently \citet{2019arXiv190901368W} report that $z\sim0$ galaxies with weak {\sc[Sii]} emission are Lyman continuum leakers.
Since the ionization potential of {\sc Sii} is 10.4 eV, less than that of {\sc Hii} (13.6 eV), most of the {\sc[Sii]} emission come from PDRs.
Thus weak {\sc[Sii]} emission would be signposts of the low PDR covering fraction, resulting the Lyman continuum leakage.
Because the ionization potential of {\sc Cii} is 11.3 eV, less than 13.6 eV, the weak {\sc[Cii]} emission will be also signposts of the Lyman continuum leakage.
Therefore, the $z=6-9$ galaxies with low $L_\m{[CII]}/SFR$ (and high $L_\m{[OIII]}/L_\m{[CII]}$) would be Lyman continuum leakers, significantly contributing to reionization.

\section{Summary}\label{ss_summary}
In this paper, we present our new ALMA observations targeting three LBGs at $z\sim6$, J1211-0118, J0235-0532, and J0217-0208, which are identified in the Subaru/HSC survey, and are already spectroscopically confirmed with Ly$\alpha$.
In conjunction with the previous ALMA observations for $z>5$ galaxies, we study {\sc[Oiii]}88$\mu$m, {\sc[Cii]}158$\mu$m, {\sc[Nii]}122$\mu$m, and dust continuum emission, and examine their relations with SFRs and rest-UV properties such as the UV slope $\beta$ and the Ly$\alpha$ EWs.
We then discuss the physical origins of the observed properties based on the model calculations with Cloudy.
Our major findings are summarized as below.

\begin{enumerate}

\item
We detect {\sc[Oiii]} and {\sc[Cii]} emission lines from all of our targets at the $4.3-11.8\sigma$ significance levels. 
The redshifts derived from the {\sc[Oiii]} and {\sc[Cii]} lines are consistent within $1\sigma$ uncertainties, and Ly$\alpha$ emission lines are redshifted from the {\sc[Oiii]} and {\sc[Cii]} redshifts by $\Delta v_\m{Ly\alpha}\simeq40-200\ \m{km\ s^{-1}}$.
The {\sc [Nii]} emission are not detected at $>3\sigma$ in our targets.

\item 
The {\sc[Oiii]} luminosities of our targets are comparable to or higher than the $L_\m{[OIII]}-SFR$ relation of $z\sim0$ low metallicity dwarf galaxies.
The {\sc[Cii]} luminosities are lower than those of the dwarf galaxies, and comparable to the local starburst galaxies.
As a result, our galaxies show $\sim10$ times higher {\sc[Oiii]}/{\sc[Cii]} ratios than local galaxies, similar to other $z=6-9$ galaxies in the literature.

\item 
In J1211-0118 and J0217-0208, the {\sc[Cii]} emission are spatially well resolved, and show velocity gradients of $\Delta v_\m{obs}\sim220-250\ \m{km\ s^{-1}}$.
The ratios of the velocity gradient to the velocity dispersion are $\Delta v_\m{obs}/2\sigma_\m{tot}>0.4$, indicating that J1211-0118 and J0217-0208 are consistent with rotation supported systems.

\item 
We identify dust continuum emission at $120\ \m{\mu m}$ and $160\ \m{\mu m}$ in J1211-0118 and J0217-0208, but not in J0235-0532.
We fit the observed continuum fluxes and upper limits with the modified black body, and obtain the IR luminosities of $L_\m{IR}\sim10^{11}\ L_\odot$, and the dust temperatures of $T_\m{dust}\sim30-40\ \m{K}$ for J1211-0118 and J0217-0208.
We also obtain the upper limit of $L_\m{IR}<2.5\times10^{11}\ L_\odot$ with $T_\m{dust}<50\ \m{K}$ for J0235-0532.

\item
J1211-0118 follows the Calzetti IRX-$\beta_\m{UV}$ relation, and the upper limit of J0235-0532 is consistent with both the Calzetti, Takeuchi, and SMC relations.
The IRX value of J0217-0208 is lower than the SMC IRX-$\beta$ relation, which would be due to the geometry effect, AGN activity, or the high molecular gas fraction.

\item 
Based on the Cloudy calculations, we discuss the physical origins of the high {\sc[Oiii]}/{\sc[Cii]} ratios and low $L_\m{[CII]}/SFR$ ratios of the $z=6-9$ galaxies compared to the $z\sim0$ galaxies.
We find that the properties of the $z=6-9$ galaxies can be explained by $\times10-100$ higher ionization parameters or low PDR covering fractions of $0-10\%$ like a density bounded nebula, possibly due to the young stellar populations, compact sizes of the $z=6-9$ galaxies, or outflow, consistent with our {\sc[Nii]} observations.
The low PDR covering fraction would enhancing the Ly$\alpha$, Lyman continuum, and $\mathrm{C^+}$ ionizing photon escapes from galaxies.
Higher hydrogen density, lower $\m{C/O}$ ratio, and the CMB attenuation effect can reproduce a part of the $z=6-9$ galaxies, but not all of the $z=6-9$ galaxies with {\sc [Oiii]} and {\sc [Cii]} observations by itself, and the combination of these effects is needed.

\item 
Including our new observations, we find the anti-correlation between $L_\m{[CII]}/SFR$ and $EW^{0,\m{int}}_\m{Ly\alpha}$ at the $>95\%$ significance level.
We also identify the correlation between the {\sc[Oiii]}/{\sc[Cii]} ratio and the Ly$\alpha$ EW at the $\sim90\%$ significance level.
These relations indicate that the origin of the {\sc[Cii]} deficit in LAEs would be the high ionization parameters or low PDR covering fractions, which make the Ly$\alpha$ and Lyman continuum photons escape easier.

\end{enumerate}

\acknowledgments
We thank the anonymous referee for a careful reading and valuable comments that improved the clarity of the paper.
We thank Daisuke Iono, Thomas Greve, Chris Matzner, Tomonari Michiyama, Kentaro Nagamine, Daniel Schaerer, Masayuki Umemura, and Hidenobu Yajima for their useful comments and discussions.

This paper makes use of the following ALMA data: ADS/JAO.ALMA \#2017.1.00508.S.
ALMA is a partnership of ESO (representing its member states), NSF (USA) and NINS (Japan), together with NRC (Canada), MOST and ASIAA (Taiwan), and KASI (Republic of Korea), in cooperation with the Republic of Chile.
The Joint ALMA Observatory is operated by ESO, AUI/NRAO and NAOJ.

The Hyper Suprime-Cam (HSC) collaboration includes the astronomical communities of Japan and Taiwan, and Princeton University.  The HSC instrumentation and software were developed by the National Astronomical Observatory of Japan (NAOJ), the Kavli Institute for the Physics and Mathematics of the Universe (Kavli IPMU), the University of Tokyo, the High Energy Accelerator Research Organization (KEK), the Academia Sinica Institute for Astronomy and Astrophysics in Taiwan (ASIAA), and Princeton University.  Funding was contributed by the FIRST program from Japanese Cabinet Office, the Ministry of Education, Culture, Sports, Science and Technology (MEXT), the Japan Society for the Promotion of Science (JSPS),  Japan Science and Technology Agency  (JST),  the Toray Science  Foundation, NAOJ, Kavli IPMU, KEK, ASIAA,  and Princeton University.

The Pan-STARRS1 Surveys (PS1) have been made possible through contributions of the Institute for Astronomy, the University of Hawaii, the Pan-STARRS Project Office, the Max-Planck Society and its participating institutes, the Max Planck Institute for Astronomy, Heidelberg and the Max Planck Institute for Extraterrestrial Physics, Garching, The Johns Hopkins University, Durham University, the University of Edinburgh, Queen's University Belfast, the Harvard-Smithsonian Center for Astrophysics, the Las Cumbres Observatory Global Telescope Network Incorporated, the National Central University of Taiwan, the Space Telescope Science Institute, the National Aeronautics and Space Administration under Grant No. NNX08AR22G issued through the Planetary Science Division of the NASA Science Mission Directorate, the National Science Foundation under Grant No. AST-1238877, the University of Maryland, and Eotvos Lorand University (ELTE).

This paper makes use of software developed for the Large Synoptic Survey Telescope. We thank the LSST Project for making their code available as free software at http://dm.lsst.org.

This work is supported by World Premier International Research Center Initiative (WPI Initiative), MEXT, Japan.
Y. Harikane acknowledges support from the Advanced Leading Graduate Course for Photon Science (ALPS) grant and the JSPS KAKENHI grant No. 16J03329 and 19J01222 through the JSPS Research Fellowship for Young Scientists.
A.K. Inoue was supported by the JSPS KAKENHI grant No. 17H01114 and by the NAOJ ALMA grant No. 2016-01A.
Y. Matsuoka was supported by the JSPS KAKENHI grant No. 17H04830 and the Mitsubishi Foundation
grant No. 30140.
T. Nagao was supported by the JSPS KAKENHI grant No. 16H03958 and 19H00697.
Y. Matsuda acknowledges support from the JSPS grants 17H04831, 17KK0098 and 19H00697.
L. Vallini acknowledges funding from the European Union’s Horizon 2020 research and innovation program under the Marie Sklodowska-Curie Grant agreement No. 746119.

\begin{deluxetable*}{ccccccc}
\setlength{\tabcolsep}{0.4cm}
\tablecaption{Summary of High Redshift Galaxies with ALMA Observations}
\tablehead{\colhead{Name} & \colhead{$z_\m{spec}$} & \colhead{$L_\m{[CII]}$}  & \colhead{$L_\m{[OIII]}$} & \colhead{$SFR_\m{tot}$} & \colhead{$EW_\m{Ly\alpha}^0$} & \colhead{Ref.} \\
\colhead{(1)}& \colhead{(2)}& \colhead{(3)}& \colhead{(4)} &  \colhead{(5)}& \colhead{(6)}& \colhead{(7)}}
\startdata
\multicolumn{7}{c}{LBGs \& LAEs}\\
MACS1149-JD1 & 9.110 & $<4.0\times10^{6}$ & $(7.4\pm1.6)\times10^{7}$ & $4.2^{+0.8}_{-1.1}$ & $10$ & H18a, L19 \\
A2744-YD4 & 8.382 & $<2.0\times10^{7}$ & $(7.0\pm1.7)\times10^{7}$ & $12.9^{+11.1}_{-6.0}$ & $10.7\pm2.7$ & L17a,19 \\
MACS0416-Y1 & 8.312 & $(1.4\pm0.2)\times10^{8}$ & $(1.2\pm0.3)\times10^{9}$ & $57.0^{+175.0}_{-0.2}$ & $<16.7$ & Tam19, B20 \\
SXDF-NB1006-2 & 7.215 & $<8.4\times10^{7}$ & $(9.9\pm2.1)\times10^{8}$ & $219^{+105}_{-176}$ & $>15.4$ & I16 \\
B14-65666 & 7.168 & $(1.1\pm0.1)\times10^{9}$ & $(3.4\pm0.4)\times10^{9}$ & $200^{+82}_{-38}$ & $3.7_{-1.1}^{+1.7}$ & H19, F16 \\
BDF-3299 & 7.109 & $(4.9\pm0.6)\times10^{7}$ & $(1.8\pm0.2)\times10^{8}$ & $5.4$ & $38.8$ & C17, M15 \\
J0217-0208 & 6.204 & $(1.4\pm0.2)\times10^{9}$ & $(8.5\pm2.0)\times10^{9}$ & $96$ & $15\pm1$ & This work \\
J0235-0532 &6.090 & $(4.3\pm0.7)\times10^{8}$ & $(3.8\pm0.3)\times10^{9}$ & $54$ & $41\pm2$ & This work \\
J1211-0118 & 6.029 & $(1.4\pm0.1)\times10^{9}$ & $(4.8\pm0.7)\times10^{9}$ & $86$ & $6.9\pm0.8$ & This work \\
z8-GND-5296 & 7.508 & $<3.5\times10^{8}$ & $\dots$ & $14.7$ & $8$ & S15, F12 \\
A1689-zD1 & 7.5 & $<8.9\times10^{7}$ & $\dots$ & $11.7^{+4.1}_{-2.2}$ & $<27$ & W15 \\
COSMOS13679 & 7.154 & $(7.4\pm1.7)\times10^{7}$ & $\dots$ & $15.1$ & $15$ & P16 \\
BDF-521 & 7.109 & $<6.0\times10^{7}$ & $\dots$ & $5.6$ & $64$ & M15, V11 \\
IOK-1 & 6.965 & $<3.4\times10^{7}$ & $\dots$ & $15.1\pm0.9$ & $43$ & O14, O12 \\
COS-3018555981 & 6.854 & $(4.7\pm0.5)\times10^{8}$ & $\dots$ & $19.2\pm1.6$ & $<2.9$ & S17, L17 \\
SDF46975 & 6.844 & $<5.8\times10^{7}$ & $\dots$ & $14.5$ & $43$ & M15, O12 \\
COS-2987030247 & 6.816 & $(3.6\pm0.5)\times10^{8}$ & $\dots$ & $22.7\pm2.0$ & $16.2^{+5.2}_{-5.5}$ & S17, L17 \\
RXJ1347-1145 & 6.765 & $(1.5^{+0.2}_{-0.4})\times10^{7}$ & $\dots$ & $8.5^{+5.9}_{-1.0}$ & $26\pm4$ & B16 \\
NTTDF6345 & 6.701 & $(1.9\pm0.3)\times10^{8}$ & $\dots$ & $9.4$ & $15$ & P16 \\
UDS16291 & 6.638 & $(7.2\pm1.7)\times10^{7}$ & $\dots$ & $10.0$ & $6$ & P16 \\
COSMOS24108 & 6.629 & $(1.0\pm0.2)\times10^{8}$ & $\dots$ & $18.3$ & $27$ & P16 \\
CR7 & 6.604 & $(2.0\pm0.4)\times10^{8}$ & $\dots$ & $28.4\pm1.3$ & $211\pm20$ & M17, So15 \\
Himiko & 6.595 & $(1.2\pm0.2)\times10^{8}$ & $\dots$ & $19.2\pm1.3$ & $78^{+8}_{-6}$ & C18, O13 \\
UDS4821 & 6.561 & $<6.8\times10^{7}$ & $\dots$ & $13$ & $48$ & C17 \\
HCM6A & 6.56 & $<6.5\times10^{7}$ & $\dots$ & $6.3$ & $25.1$ & K13, H02 \\
MASOSA & 6.543 & $<6.6\times10^{7}$ & $\dots$ & $9.5_{-1.3}^{+2.5}$ & $145^{+50}_{-43}$ & M19\\
VR7 & 6.529 & $(4.8\pm0.4)\times10^{8}$ & $\dots$ & $34.0_{-1.3}^{+3.2}$ & $34^{+4}_{-4}$ & M19\\
COSMOS20521 & 6.36 & $<4.8\times10^{7}$ & $\dots$ & $14$ & $10$ & C17 \\
GOODS3203 & 6.27 & $<1.2\times10^{8}$ & $\dots$ & $18$ & $5$ & C17 \\
CLM1 & 6.176 & $(2.4\pm0.3)\times10^{8}$ & $\dots$ & $37\pm4$ & $50$ & W15, C03 \\
BDF2203 & 6.12 & $(1.3\pm0.3)\times10^{8}$ & $\dots$ & $16$ & $3$ & C17 \\
WMH5 & 6.076 & $(6.6\pm0.8)\times10^{8}$ & $\dots$ & $43\pm5$ & $13\pm4$ & W15, W13 \\
NTTDF2313 & 6.07 & $<4.5\times10^{7}$ & $\dots$ & $12$ & $0$ & C17 \\
A383-5.1 & 6.029 & $(8.9\pm3.1)\times10^{6}$ & $\dots$ & $2.0$ & $138$ & K16, St15 \\
WMH13 & 5.985 & $(1.1\pm0.2)\times10^{8}$ & $\dots$ & $24.9$ & $27$ & F19 \\
HZ1 & 5.690 & $(2.5\pm1.9)\times10^{8}$ & $\dots$ & $24^{+6}_{-3}$ & $5.3^{+2.6}_{-4.1}$ & C15, M12 \\
NB816-S-61269 & 5.684 & $(2.1\pm0.5)\times10^{8}$ & $\dots$ & $14.4$ & $93.3$ & F19 \\
HZ2 & 5.670 & $(3.6\pm3.4)\times10^{8}$ & $\dots$ & $25^{+5}_{-2}$ & $6.9\pm2.0$ & C15 \\
HZ10 & 5.659 & $(1.3\pm0.4)\times10^{9}$ & $\dots$ & $169^{+32}_{-27}$ & $24.5^{+9.2}_{-11.0}$ & C15, M12 \\
HZ9 & 5.548 & $(1.6\pm0.3)\times10^{9}$ & $\dots$ & $67^{+30}_{-20}$ & $14.4^{+6.8}_{-5.4}$ & C15, M12 \\
HZ3 & 5.546 & $(4.7\pm3.0)\times10^{8}$ & $\dots$ & $18^{+8}_{-3}$ & $<3.6$ & C15 \\
HZ4 & 5.540 & $(9.5\pm4.8)\times10^{8}$ & $\dots$ & $51^{+54}_{-18}$ & $10.2^{+0.9}_{-4.4}$ & C15, M12 \\
HZ6 & 5.290 & $(1.4\pm0.6)\times10^{9}$ & $\dots$ & $49^{+44}_{-12}$ & $8.0^{+12.1}_{-7.3}$ & C15, M12 \\
HZ7 & 5.250 & $(5.5\pm3.0)\times10^{8}$ & $\dots$ & $21^{+5}_{-2}$ & $9.8\pm5.5$ & C15 \\
HZ8 & 5.148 & $(2.6\pm1.1)\times10^{8}$ & $\dots$ & $18^{+5}_{-2}$ & $27.1^{+12.9}_{-14.7}$ & C15, M12 \\
\hline
\multicolumn{7}{c}{SMGs}\\
SPT0311-58-E & 6.900 & $(5.4\pm0.5)\times10^{9}$ & $(6.9\pm0.7)\times10^{9}$ & $540\pm175$ & $\dots$ & M18 \\
SPT0311-58-W & 6.900 & $(1.0\pm0.2)\times10^{10}$ & $(5.7\pm1.1)\times10^{9}$ & $2900\pm1800$ & $\dots$ & M18 \\
J2100-SB & 6.0806 & $(1.8\pm0.9)\times10^{9}$ & $(2.9\pm0.1)\times10^{9}$ & $284\pm30$ & $\dots$ & W18 \\
COSMOS-AzTEC-1 & 4.342 & $(6.3\pm0.6)\times10^{9}$ & $(2.2\pm0.7)\times10^{9}$ & $1169^{+11}_{-274}$ & $\dots$ & Tad18,19
\enddata
\tablecomments{(1) Object Name.
(2) Redshift determined with Ly$\alpha$, Lyman break, rest-frame UV absorption lines, {[\sc Cii]}158$\m{\mu m}$, or {[\sc Oiii]}88$\m{\mu m}$.
(3) {[\sc Cii]}158$\m{\mu m}$ luminosity or its $3\sigma$ upper limit in units of $L_\odot$.
(4) {[\sc Oiii]}88$\m{\mu m}$ luminosity in units of $L_\odot$.
(5) Total SFR ($=SFR_\m{UV}+SFR_\m{IR}$) in units of $M_\odot\ \m{yr^{-1}}$.
(6) Rest-frame Ly$\alpha$ EW not corrected for the inter-galactic medium (IGM) absorption in units of $\m{\AA}$.
(7) Reference (B17: \citealt{2017ApJ...836L...2B}, 
B20: \citealt{2020MNRAS.493.4294B},
C03: \citealt{2003A&A...405L..19C}, 
C15: \citealt{2015Natur.522..455C}, 
C17: \citealt{2017A&A...605A..42C},
C18a: \citealt{2018ApJ...854L...7C},
C18b: \citealt{2018MNRAS.478.1170C}, 
F13: \citealt{2013Natur.502..524F},
F16: \citealt{2016ApJ...822...46F},
F19: \citealt{2019arXiv190206760F},
H02: \citealt{2002ApJ...568L..75H}, 
H18a: \citealt{2018Natur.557..392H}
H19: \citealt{2019PASJ...71...71H}
I16: \citealt{2016Sci...352.1559I}, 
K13: \citealt{2013ApJ...771L..20K}, 
K16: \citealt{2016MNRAS.462L...6K}, 
L17a: \citealt{2017ApJ...837L..21L},
L17b: \citealt{2017ApJ...851...40L}, 
L19: \citealt{2019MNRAS.487L..81L}
M12: \citealt{2012ApJ...760..128M},
M15: \citealt{2015MNRAS.452...54M}, 
M17: \citealt{2017ApJ...851..145M}, 
M18: \citealt{2018Natur.553...51M},
M19: \citealt{2019ApJ...881..124M},
O12: \citealt{2012ApJ...744...83O}, 
O13: \citealt{2013ApJ...778..102O}, 
O14: \citealt{2014ApJ...792...34O}, 
P16: \citealt{2016ApJ...829L..11P}, 
S12: \citealt{2012ApJ...752..114S}, 
Sc15: \citealt{2015A&A...574A..19S}, 
So15: \citealt{2015ApJ...808..139S}, 
St15: \citealt{2015MNRAS.450.1846S},
S17: \citealt{2017arXiv170604614S},  
Tad18: \citealt{2018Natur.560..613T},
Tad19: \citealt{2019ApJ...876....1T},
Tam19: \citealt{2019ApJ...874...27T},
V11: \citealt{2011ApJ...730L..35V}, 
W13: \citealt{2013AJ....145....4W}, 
Wa15: \citealt{2015Natur.519..327W}, 
Wi15: \citealt{2015ApJ...807..180W},
W18: \citealt{2018ApJ...869L..22W}).}
\label{tab_lite}
\end{deluxetable*}

\bibliographystyle{apj}
\bibliography{apj-jour,reference}


\end{document}